\documentclass[final,1p,times]{elsarticle}
\usepackage{amssymb}
\usepackage{amsfonts}
\usepackage{amsmath}
\usepackage{geometry}
\usepackage{bm}
\newtheorem{thm}{Theorem}

\newtheorem{defn}{Definition}
\newdefinition{rmk}{Remark}
\newproof{pf}{Proof}
\newproof{pot}{Proof of Theorem}
\begin{document}

\begin{frontmatter}
\title{The integral formalism and the generating function of grand confluent hypergeometric function}

\author{Yoon Seok Choun
}
\ead{ychoun@gc.cuny.edu; Yoon.Choun@baruh.cuny.edu; ychoun@gmail.com}
\address{Baruch College, The City University of New York, Natural Science Department, A506, 17 Lexington Avenue, New York, NY 10010}
\begin{abstract}

Biconfluent Heun (BCH) function, a confluent form of Heun function\cite{Heun1889,Ronv1995}, is the special case of Grand Confluent Hypergeometric (GCH) function\footnote{For the canonical form of BCH equation \cite{Ronv1995}, replace $\mu $, $\varepsilon $, $\nu $, $\Omega $ and $\omega $ by $-2$, $-\beta  $, $ 1+\alpha $, $\gamma -\alpha -2 $ and $ 1/2 (\delta /\beta +1+\alpha )$ in (\ref{eq:1a}). For DLFM version \cite{NIST} or in ref.\cite{Slavy2000}, replace $\mu $ and $\omega $ by 1 and $-q/\varepsilon $ in (\ref{eq:1a}).}: this has a regular singularity at $x=0$, and an irregular singularity at $\infty$ of rank 2. 

In this paper I apply three term recurrence formula (3TRF) \cite{Chou2012b} to the integral formalism of GCH function including all higher terms of $A_n$'s and the generating function for the GCH polynomial which makes $B_n$ term terminated. I show how to transform power series expansion in closed forms of GCH equation to its integral representation analytically. 

This paper is 10th out of 10 in series ``Special functions and three term recurrence formula (3TRF)''. See section 6 for all the papers in the series.  The previous paper in the series describes the power series expansion in closed forms of GCH equation and its asymtotic behaviours\cite{Chou2012i}.
\end{abstract}

                                      
\end{frontmatter}
\section{\label{sec:level1}Introduction}
\begin{equation}
x \frac{d^2{y}}{d{x}^2} + \left( \mu x^2 + \varepsilon x + \nu  \right) \frac{d{y}}{d{x}} + \left( \Omega x + \varepsilon \omega \right) y = 0
\label{eq:1a}
\end{equation}
(\ref{eq:1a}) is  Grand Confluent Hypergeometric (GCH) differential equation where $\mu$, $\varepsilon$, $\nu $, $\Omega$ and $\omega$ are real or imaginary parameters \cite{Chou2012a,Chou2012i}.
GCH ordinary differential equation is of Fuchsian types with the one regular and one irregular singularities. In contrast, Heun equation of Fuchsian types has the four regular singularities. Heun equation has the four kind of confluent forms: (1) Confluent Heun (two regular and one irregular singularities), (2) Doubly confluent Heun (two irregular singularities), (3) Biconfluent Heun (one regular and one irregular singularities), (4) Triconfluent Heun equations (one irregular singularity). BCH equation is derived from the GCH equation by changing all coefficients.\cite{Ronv1995}

In previous paper I construct analytic solutions of GCH function for all higher terms of $A_n$'s \cite{Chou2012i} by applying three term recurrence formula.\cite{Chou2012b}; for power series expansions of infinite series and polynomial, its asymptotic behaviors and boundary conditions for an independent variable $x$.

In this paper I consider an integral form of GCH function and the generating function for the GCH polynomial which makes $B_n$ term terminated. Since the integral form of GCH function is constructed, the GCH function is able to be transformed to other well-known special functions analytically such as Bessel, Kummer and Hypergeometric functions, etc.

I already obtained approximative normalized wave function of the spin free Hamiltonian involving only scalar potential for the $q-\bar{q}$ system up to first order of extremely small mass of quark\cite{Chou2012a}. According to this paper we might be possible to obtain the analytic normalized wave function from the generating function for the GCH polynomial. 

This new wave function has infinite eigenvalues \cite{Chou2012a}. Because a GCH differential equation consists of three recursive coefficients\cite{Chou2012b}. In contrast any differential equations having two recursive coefficients have only one eigenvalue; i.e. the wave function for hydrogen-like atom. 

We can apply GCH function into modern physics \cite{Slav1996,Ralk2002,Kand2005,Hortacsu:2011rr,Arri1991}. Section 4 contain three additional examples using power series expansion in closed forms and its integral form of GCH function. 
\section{\label{sec:level2}Integral formalism}

\subsection{Polynomial which makes $B_n$ term terminated}
In this article Pochhammer symbol $(x)_n$ is used to represent the rising factorial: $(x)_n = \frac{\Gamma (x+n)}{\Gamma (x)}$.
There is a generalized hypergeometric function which is written by
\begin{eqnarray}
I_j &=& \sum_{i_j= i_{j-1}}^{\beta _j} \frac{(-\beta _j)_{i_j}(1+\frac{j}{2}+\frac{\lambda }{2})_{i_{j-1}}(\frac{j}{2}+\gamma +\frac{\lambda }{2})_{i_{j-1}}}{(-\beta _j)_{i_{j-1}}(1+\frac{j}{2}+\frac{\lambda }{2})_{i_j}(\frac{j}{2}+\gamma +\frac{\lambda }{2})_{i_j}} z^{i_j}\nonumber\\
&=& z^{i_{j-1}} 
\sum_{l=0}^{\infty } \frac{B(i_{j-1}+\frac{j}{2}+\frac{\lambda }{2},l+1) B(i_{j-1}-1+\gamma +\frac{j}{2}+\frac{\lambda }{2},l+1)(-\beta _j+i_{j-1})_l}{(i_{j-1}+\frac{j}{2}+\frac{\lambda }{2})^{-1}(i_{j-1}-1+\gamma +\frac{j}{2}+ \frac{\lambda }{2})^{-1}(1)_l \;l!} z^l 
\label{eq:1}
\end{eqnarray}
By using integral form of beta function,
\begin{subequations}
\begin{equation}
B\left(i_{j-1}+\frac{j}{2}+\frac{\lambda }{2}, l+1\right)= \int_{0}^{1} dt_j\;t_j^{i_{j-1}+\frac{j}{2}-1+\frac{\lambda }{2}} (1-t_j)^l
\label{eq:2a}
\end{equation}
\begin{equation}
 B\left(i_{j-1}+\gamma -1+\frac{j}{2}+\frac{\lambda }{2}, l+1\right)= \int_{0}^{1} du_j\;u_j^{i_{j-1}+\gamma -2+\frac{j}{2}+\frac{\lambda }{2}} (1-u_j)^l
 \label{eq:2b}
\end{equation}
\end{subequations}
Substitute (\ref{eq:2a}) and (\ref{eq:2b}) into (\ref{eq:1}), and divide $(i_{j-1}+\frac{j}{2}+\frac{\lambda }{2})(i_{j-1}-1+\gamma +\frac{j}{2}+ \frac{\lambda }{2})$ into the new (\ref{eq:1}).
\begin{eqnarray}
&& \frac{1}{(i_{j-1}+\frac{j}{2}+\frac{\lambda }{2})(i_{j-1}-1+\gamma +\frac{j}{2}+ \frac{\lambda }{2})}
 \sum_{i_j= i_{j-1}}^{\beta _j} \frac{(-\beta _j)_{i_j}(1+\frac{j}{2}+\frac{\lambda }{2})_{i_{j-1}}(\frac{j}{2}+\gamma +\frac{\lambda }{2})_{i_{j-1}}}{(-\beta _j)_{i_{j-1}}(1+\frac{j}{2}+\frac{\lambda }{2})_{i_j}(\frac{j}{2}+\gamma +\frac{\lambda }{2})_{i_j}} z^{i_j}\nonumber\\
&=&  \int_{0}^{1} dt_j\;t_j^{\frac{j}{2}-1+\frac{\lambda }{2}} \int_{0}^{1} du_j\;u_j^{\gamma -2+\frac{j}{2}+\frac{\lambda }{2}} (z t_j u_j)^{i_{j-1}}
 \sum_{l=0}^{\infty } \frac{(-(\beta _j-i_{j-1}))_l}{(1)_l \;l!} [z(1-t_j)(1-u_j)]^l
 \label{eq:3}
\end{eqnarray}
Confluent hypergeometric polynomial of the first kind is defined by
\begin{equation}
F_{\beta _0}(\gamma ;z)= \frac{\Gamma (\beta _0+\gamma )}{\Gamma (\gamma )} \sum_{n=0}^{\infty } \frac{(-\beta _0)_n}{(\gamma )_n\;n!} z^n = \frac{\beta _0!}{2\pi i}  \oint dv \frac{\exp\left(-\frac{zv}{(1-v)}\right)}{v^{\beta _0+1}(1-v)^{\gamma }}
 \label{eq:4}
\end{equation}
Replace $\beta _0$, $\gamma $, $v$ and $z$ by $\beta _j-i_{j-1}$, 1, $v_j$ and $z(1-t_j)(1-u_j)$ in (\ref{eq:4}), and divide $\Gamma (\beta _j+1-i_{j-1})$ on the new (\ref{eq:4}).
\begin{eqnarray}
\frac{F_{\beta _j-i_{j-1}}\left(\gamma =1;z(1-t_j)(1-u_j)\right)}{\Gamma (\beta _j+1-i_{j-1})}&=& \frac{1}{2\pi i}  \oint dv_j \frac{\exp\left(-\frac{v_j}{(1-v_j)}z(1-t_j)(1-u_j)\right)}{v_j^{\beta _j+1-i_{j-1}}(1-v_j)}\nonumber\\
&=& \sum_{l=0}^{\infty } \frac{(-(\beta _j-i_{j-1}))_l}{(1)_l\;l!} [z(1-t_j)(1-u_j)]^l
 \label{eq:5}
\end{eqnarray}
Substitute (\ref{eq:5}) into (\ref{eq:3}).
\begin{eqnarray}
K_j &=& \frac{1}{(i_{j-1}+\frac{j}{2}+\frac{\lambda }{2})(i_{j-1}-1+\gamma +\frac{j}{2}+ \frac{\lambda }{2})}
\sum_{i_j= i_{j-1}}^{\beta _j} \frac{(-\beta _j)_{i_j}(1+\frac{j}{2}+\frac{\lambda }{2})_{i_{j-1}}(\frac{j}{2}+\gamma +\frac{\lambda }{2})_{i_{j-1}}}{(-\beta _j)_{i_{j-1}}(1+\frac{j}{2}+\frac{\lambda }{2})_{i_j}(\frac{j}{2}+\gamma +\frac{\lambda }{2})_{i_j}} z^{i_j} \nonumber\\
&=&  \int_{0}^{1} dt_j\;t_j^{\frac{j}{2}-1+\frac{\lambda }{2}} \int_{0}^{1} du_j\;u_j^{\gamma -2+\frac{j}{2}+\frac{\lambda }{2}} 
\frac{1}{2\pi i}  \oint dv_j \frac{\exp\left(-\frac{v_j}{(1-v_j)}z(1-t_j)(1-u_j)\right)}{v_j^{\beta _j+1}(1-v_j)}(z t_j u_j v_j)^{i_{j-1}}\hspace{1cm} \label{eq:6}
\end{eqnarray}
In Ref.\cite{Chou2012i} the general expression of power series of GCH equation for polynomial which makes $B_n$ term terminated is given by; $\lambda $ is indicial roots which are 0 or $1-\nu $
\begin{eqnarray}
 y(x)&=& \sum_{n=0}^{\infty } y_{n}(x) = y_0(x)+ y_1(x)+ y_2(x)+y_3(x)+\cdots \nonumber\\
&=& c_0 x^{\lambda } \Bigg\{\sum_{i_0=0}^{\beta _0} \frac{(-\beta _0)_{i_0}}{(1+\frac{\lambda }{2})_{i_0}(\gamma +\frac{\lambda }{2})_{i_0}}z^{i_0}
+   \Bigg\{ \sum_{i_0=0}^{\beta _0}\frac{(i_0+\frac{\lambda }{2}+\frac{\omega }{2})}{(i_0+\frac{1}{2}+\frac{\lambda }{2})(i_0-\frac{1}{2}+\gamma +\frac{\lambda }{2})} \frac{(-\beta _0)_{i_0}}{(1+\frac{\lambda }{2})_{i_0}(\gamma +\frac{\lambda }{2})_{i_0}}\nonumber\\
&&\times  \sum_{i_1=i_0}^{\beta _1} \frac{(-\beta _1)_{i_1}(\frac{3}{2}+\frac{\lambda }{2})_{i_0}(\gamma +\frac{1}{2}+ \frac{\lambda }{2})_{i_0}}{(-\beta _1)_{i_0}(\frac{3}{2}+\frac{\lambda }{2})_{i_1}(\gamma +\frac{1}{2}+\frac{\lambda }{2})_{i_1}} z^{i_1} \Bigg\}\tilde{\varepsilon } \nonumber\\
&&+ \sum_{N=2}^{\infty } \Bigg\{ \sum_{i_0=0}^{\beta _0} \frac{(i_0+\frac{\lambda }{2}+\frac{\omega }{2})}{(i_0+\frac{1}{2}+\frac{\lambda }{2})(i_0-\frac{1}{2}+\gamma +\frac{\lambda }{2})} \frac{(-\beta _0)_{i_0}}{(1+\frac{\lambda }{2})_{i_0}(\gamma +\frac{\lambda }{2})_{i_0}} \nonumber\\
&&\times \prod _{k=1}^{n-1} \Bigg\{ \sum_{i_k=i_{k-1}}^{\beta _k} \frac{(i_k+\frac{\lambda }{2}+\frac{\omega }{2}+\frac{k}{2})}{(i_k+\frac{1}{2}+\frac{\lambda }{2}+\frac{k}{2})(i_k-\frac{1}{2}+\gamma + \frac{k}{2}+\frac{\lambda }{2})} 
 \frac{(-\beta _k)_{i_k}(1+\frac{k}{2}+\frac{\lambda }{2})_{i_{k-1}}(\frac{k}{2}+\gamma +\frac{\lambda }{2})_{i_{k-1}}}{(-\beta _k)_{i_{k-1}}(1+\frac{k}{2}+\frac{\lambda }{2})_{i_k}(\frac{k}{2}+\gamma +\frac{\lambda }{2})_{i_k}}\Bigg\} \nonumber\\
&&\times \sum_{i_n= i_{n-1}}^{\beta _n} \frac{(-\beta _n)_{i_n}(1+\frac{n}{2}+\frac{\lambda }{2})_{i_{n-1}}(\frac{n}{2}+\gamma +\frac{\lambda }{2})_{i_{n-1}}}{(-\beta _n)_{i_{n-1}}(1+\frac{n}{2}+\frac{\lambda }{2})_{i_n}(\frac{n}{2}+\gamma +\frac{\lambda }{2})_{i_n}} z^{i_n}\Bigg\} \tilde{\varepsilon }^n \Bigg\}
\label{eq:7}
\end{eqnarray}
where
\begin{equation}
\begin{cases} z = -\frac{1}{2}\mu x^2 \cr
\tilde{\varepsilon }  = -\frac{1}{2}\varepsilon  x\cr
\gamma  = \frac{1}{2}(1+\nu ) \cr
\Omega = -\mu (2\beta _i+i+\lambda )\;\;\mbox{as}\;i=0,1,2,\cdots \;\;\mbox{and}\;\; \beta_i= 0,1,2,\cdots \cr
\mbox{As}\; i\leq j\rightarrow \beta _i\leq \beta _j 
\end{cases}
\nonumber
\end{equation}
Substitute (\ref{eq:6}) into (\ref{eq:7}) where $j=1,2,3,\cdots$; apply $K_1$ into the second summation of sub-power series $y_1(x)$, apply $K_2$ into the third summation and $K_1$ into the second summation of sub-power series $y_2(x)$, apply $K_3$ into the forth summation, $K_2$ into the third summation and $K_1$ into the second summation of sub-power series $y_3(x)$, etc.\footnote{$y_1(x)$ means the sub-power series in (\ref{eq:7}) contains one term of $A_n's$, $y_2(x)$ means the sub-power series in (\ref{eq:7}) contains two terms of $A_n's$, $y_3(x)$ means the sub-power series in (\ref{eq:7}) contains three terms of $A_n's$, etc.}
\begin{thm}
 The general expression of the integral representation of the GCH polynomial which makes $B_n$ term terminated is given by
\begin{eqnarray}
 y(x)&=& \sum_{n=0}^{\infty } y_{n}(x) =y_0(x)+ y_1(x)+ y_2(x)+y_3(x)+\cdots \nonumber\\
&=& c_0 x^{\lambda } \left\{ \sum_{i_0=0}^{\beta _0 }\frac{(-\beta _0)_{i_0}}{(1+\frac{\lambda }{2})_{i_0}(\gamma +\frac{\lambda }{2})_{i_0}}  z^{i_0} 
+ \sum_{n=1}^{\infty } \Bigg\{\prod _{j=0}^{n-1} \Bigg\{ \int_{0}^{1} dt_{n-j}\;t_{n-j}^{\frac{1}{2}(n-j)-1+\frac{\lambda }{2}} \int_{0}^{1} du_{n-j}\;u_{n-j}^{\gamma +\frac{1}{2}(n-j)-2+\frac{\lambda }{2}} \right.\nonumber\\
&&\times \frac{1}{2\pi i}  \oint dv_{n-j} \frac{\exp\left(-\frac{v_{n-j}}{(1-v_{n-j})}w_{n-j+1,n}(1-t_{n-j})(1-u_{n-j})\right)}{v_{n-j}^{\beta  _{n-j}+1}(1-v_{n-j})}  \left( w_{n-j,n}\partial _{w_{n-j,n}} +\frac{1}{2}\Big(n-j-1+\omega +\lambda \Big)\right) \Bigg\}\nonumber\\
&&\times \left. \sum_{i_0=0}^{\beta _0}\frac{(-\beta _0)_{i_0}}{(1+\frac{\lambda }{2})_{i_0}(\gamma +\frac{\lambda }{2})_{i_0}}  w_{1,n}^{i_0}\Bigg\}  \tilde{\varepsilon }^n \right\}
\label{eq:9}
\end{eqnarray}
where
\begin{equation}w_{a,b}=
\begin{cases} \displaystyle {z\prod _{l=a}^{b} t_l u_l v_l }\;\;\mbox{where}\; a\leq b\cr
z \;\;\mbox{only}\;\mbox{if}\; a>b
\end{cases}
\nonumber
\end{equation}
\end{thm}
\begin{pot} 
In (\ref{eq:7}) sub-power series $y_0(x) $, $y_1(x)$, $y_2(x)$ and $y_3(x)$ of the GCH polynomial which makes $B_n$ term terminated are given by
\begin{equation}
 y(x)= \sum_{n=0}^{\infty } y_{n}(x) = y_0(x)+ y_1(x)+ y_2(x)+y_3(x)+\cdots \label{eq:100}
\end{equation}
where
\begin{subequations}
\begin{equation}
 y_0(x)= c_0 x^{\lambda } \sum_{i_0=0}^{\beta _0 }\frac{(-\beta _0)_{i_0}}{(1+\frac{\lambda }{2})_{i_0}(\gamma +\frac{\lambda }{2})_{i_0}}  z^{i_0} \label{eq:101a}
\end{equation}
\begin{eqnarray}
 y_1(x)&=&  c_0 x^{\lambda } \Bigg\{ \sum_{i_0=0}^{\beta _0}\frac{(i_0+\frac{\lambda }{2}+\frac{\omega }{2})}{(i_0+\frac{1}{2}+\frac{\lambda }{2})(i_0-\frac{1}{2}+\gamma +\frac{\lambda }{2})} \frac{(-\beta _0)_{i_0}}{(1+\frac{\lambda }{2})_{i_0}(\gamma +\frac{\lambda }{2})_{i_0}}\nonumber\\
&&\times  \sum_{i_1=i_0}^{\beta _1} \frac{(-\beta _1)_{i_1}(\frac{3}{2}+\frac{\lambda }{2})_{i_0}(\gamma +\frac{1}{2}+ \frac{\lambda }{2})_{i_0}}{(-\beta _1)_{i_0}(\frac{3}{2}+\frac{\lambda }{2})_{i_1}(\gamma +\frac{1}{2}+\frac{\lambda }{2})_{i_1}} z^{i_1} \Bigg\}\tilde{\varepsilon } \label{eq:101b}
\end{eqnarray}
\begin{eqnarray}
 y_2(x) &=& c_0 x^{\lambda }\Bigg\{  \sum_{i_0=0}^{\beta _0}\frac{(i_0+\frac{\lambda }{2}+\frac{\omega }{2})}{(i_0+\frac{1}{2}+\frac{\lambda }{2})(i_0-\frac{1}{2}+\gamma +\frac{\lambda }{2})} \frac{(-\beta _0)_{i_0}}{(1+\frac{\lambda }{2})_{i_0}(\gamma +\frac{\lambda }{2})_{i_0}}\nonumber\\
&&\times  \sum_{i_1=i_0}^{\beta _1} \frac{(i_1+\frac{1}{2}+\frac{\lambda }{2}+\frac{\omega }{2})}{(i_1+1+\frac{\lambda }{2})(i_1+\gamma +\frac{\lambda }{2})} \frac{(-\beta _1)_{i_1}(\frac{3}{2}+\frac{\lambda }{2})_{i_0}(\gamma +\frac{1}{2}+ \frac{\lambda }{2})_{i_0}}{(-\beta _1)_{i_0}(\frac{3}{2}+\frac{\lambda }{2})_{i_1}(\gamma +\frac{1}{2}+\frac{\lambda }{2})_{i_1}} \nonumber\\
&&\times \sum_{i_2=i_1}^{\beta _2} \frac{(-\beta _2)_{i_2}(2+\frac{\lambda }{2})_{i_1}(\gamma +1+ \frac{\lambda }{2})_{i_1}}{(-\beta _2)_{i_1}(2+\frac{\lambda }{2})_{i_2}(\gamma +1+\frac{\lambda }{2})_{i_2}} z^{i_2} \Bigg\} \tilde{\varepsilon }^2 
\label{eq:101c}
\end{eqnarray}
\begin{eqnarray}
 y_3(x)&=&  c_0 x^{\lambda } \Bigg\{ \sum_{i_0=0}^{\beta _0}\frac{(i_0+\frac{\lambda }{2}+\frac{\omega }{2})}{(i_0+\frac{1}{2}+\frac{\lambda }{2})(i_0-\frac{1}{2}+\gamma +\frac{\lambda }{2})} \frac{(-\beta _0)_{i_0}}{(1+\frac{\lambda }{2})_{i_0}(\gamma +\frac{\lambda }{2})_{i_0}}\nonumber\\
&&\times  \sum_{i_1=i_0}^{\beta _1} \frac{(i_1+\frac{1}{2}+\frac{\lambda }{2}+\frac{\omega }{2})}{(i_1+1+\frac{\lambda }{2})(i_1+\gamma +\frac{\lambda }{2})} \frac{(-\beta _1)_{i_1}(\frac{3}{2}+\frac{\lambda }{2})_{i_0}(\gamma +\frac{1}{2}+ \frac{\lambda }{2})_{i_0}}{(-\beta _1)_{i_0}(\frac{3}{2}+\frac{\lambda }{2})_{i_1}(\gamma +\frac{1}{2}+\frac{\lambda }{2})_{i_1}} \nonumber\\
&&\times \sum_{i_2=i_1}^{\beta _2}\frac{(i_2+1+\frac{\lambda }{2}+\frac{\omega }{2})}{(i_2+\frac{3}{2}+\frac{\lambda }{2})(i_2+\frac{1}{2}+ \gamma +\frac{\lambda }{2})} \frac{(-\beta _2)_{i_2}(2+\frac{\lambda }{2})_{i_1}(\gamma +1+ \frac{\lambda }{2})_{i_1}}{(-\beta _2)_{i_1}(2+\frac{\lambda }{2})_{i_2}(\gamma +1+\frac{\lambda }{2})_{i_2}}\nonumber\\
&&\times \sum_{i_3=i_2}^{\beta _3} \frac{(-\beta _3)_{i_3}(\frac{5}{2}+\frac{\lambda }{2})_{i_2}(\gamma +\frac{3}{2}+ \frac{\lambda }{2})_{i_2}}{(-\beta _3)_{i_2}(\frac{5}{2}+\frac{\lambda }{2})_{i_3}(\gamma +\frac{3}{2}+\frac{\lambda }{2})_{i_3}} z^{i_3} \Bigg\} \tilde{\varepsilon }^3 
\label{eq:101d}
\end{eqnarray}
\end{subequations}
Put $j=1$ in (\ref{eq:6}). Take the new (\ref{eq:6}) into (\ref{eq:101b}).
\begin{eqnarray}
 y_1(x)&=& c_0 x^{\lambda }  \int_{0}^{1} dt_1\;t_1^{-\frac{1}{2}+\frac{\lambda }{2}} \int_{0}^{1} du_1\;u_1^{\gamma -\frac{3}{2}+\frac{\lambda }{2}} 
\frac{1}{2\pi i}  \oint dv_1 \frac{\exp\left(-\frac{v_1}{(1-v_1)}z(1-t_1)(1-u_1)\right)}{v_1^{\beta _1+1}(1-v_1)} \nonumber\\
&&\times \left\{ \sum_{i_0=0}^{\beta _0} \left(i_0+\frac{\omega }{2}+\frac{\lambda }{2}\right) \frac{(-\beta _0)_{i_0}}{(1+\frac{\lambda }{2})_{i_0}(\gamma +\frac{\lambda }{2})_{i_0}} (z t_1 u_1 v_1)^{i_0} \right\} \tilde{\varepsilon }\nonumber\\
&=& c_0 x^{\lambda }  \int_{0}^{1} dt_1\;t_1^{-\frac{1}{2}+\frac{\lambda }{2}} \int_{0}^{1} du_1\;u_1^{\gamma -\frac{3}{2}+\frac{\lambda }{2}} 
\frac{1}{2\pi i}  \oint dv_1 \frac{\exp\left(-\frac{v_1}{(1-v_1)}z(1-t_1)(1-u_1)\right)}{v_1^{\beta _1+1}(1-v_1)} \left( w_{1,1}\partial_{w_{1,1}} +\left( \frac{\omega }{2}+\frac{\lambda }{2}\right)\right)\nonumber\\
&&\times   \left\{ \sum_{i_0=0}^{\beta _0} \frac{(-\beta _0)_{i_0}}{(1+\frac{\lambda }{2})_{i_0}(\gamma +\frac{\lambda }{2})_{i_0}} w_{1,1}^{i_0} \right\} \tilde{\varepsilon }\label{eq:102}\\
&& \mathrm{where}\hspace{.5cm} w_{1,1}=z\prod _{l=1}^{1} t_l u_l v_l \nonumber
\end{eqnarray}
Put $j=2$ in (\ref{eq:6}). Take the new (\ref{eq:6}) into (\ref{eq:101c}). 
\begin{eqnarray}
 y_2(x)&=& c_0 x^{\lambda } \int_{0}^{1} dt_2\;t_2^{\frac{\lambda }{2}} \int_{0}^{1} du_2\;u_2^{\gamma -1+\frac{\lambda }{2}} 
\frac{1}{2\pi i}  \oint dv_2 \frac{\exp\left(-\frac{v_2}{(1-v_2)}z(1-t_2)(1-u_2)\right)}{v_2^{\beta _2+1}(1-v_2)}\left( w_{2,2}\partial _{w_{2,2}} +\Big( \frac{1}{2}+ \frac{\lambda }{2}+ \frac{\omega }{2}\Big)\right)\nonumber\\
&&\times  \Bigg\{ \sum_{i_0=0}^{\beta _0 }\frac{(i_0+\frac{\lambda }{2}+\frac{\omega }{2})}{(i_0+\frac{1}{2}+\frac{\lambda }{2})(i_0-\frac{1}{2}+\gamma +\frac{\lambda }{2})} \frac{(-\beta _0)_{i_0}}{(1+\frac{\lambda }{2})_{i_0}(\gamma +\frac{\lambda }{2})_{i_0}}\nonumber\\
&&\times  \sum_{i_1=i_0}^{\beta _1} \frac{(-\beta _1)_{i_1}(\frac{3}{2}+\frac{\lambda }{2})_{i_0}(\gamma +\frac{1}{2}+ \frac{\lambda }{2})_{i_0}}{(-\beta _1)_{i_0}(\frac{3}{2}+\frac{\lambda }{2})_{i_1}(\gamma +\frac{1}{2}+\frac{\lambda }{2})_{i_1}}w_{2,2}^{i_1} \Bigg\}\tilde{\varepsilon }^2 \label{eq:103}\\
&& \mathrm{where}\hspace{.5cm} w_{2,2}=z\prod _{l=2}^{2} t_l u_l v_l \nonumber
\end{eqnarray}
Put $j=1$ and $z =\overleftrightarrow {w}_{2,2}$ in (\ref{eq:6}). Take the new (\ref{eq:6}) into (\ref{eq:103}).
\begin{eqnarray}
 y_2(x)&=& c_0 x^{\lambda } \int_{0}^{1} dt_2\;t_2^{\frac{\lambda }{2}} \int_{0}^{1} du_2\;u_2^{\gamma -1+\frac{\lambda }{2}} 
\frac{1}{2\pi i}  \oint dv_2 \frac{\exp\left(-\frac{v_2}{(1-v_2)}z(1-t_2)(1-u_2)\right)}{v_2^{\beta _2+1}(1-v_2)} \left( w_{2,2}\partial _{w_{2,2}} +\Big( \frac{1}{2}+ \frac{\lambda }{2}+ \frac{\omega }{2}\Big)\right)\nonumber\\
&&\times  \int_{0}^{1} dt_1\;t_1^{-\frac{1}{2}+\frac{\lambda }{2}} \int_{0}^{1} du_1\;u_1^{\gamma -\frac{3}{2}+\frac{\lambda }{2}} 
\frac{1}{2\pi i}  \oint dv_1 \frac{\exp\left(-\frac{v_1}{(1-v_1)}w_{2,2}(1-t_1)(1-u_1)\right)}{v_1^{\beta _1+1}(1-v_1)} \left( w_{1,2}\partial _{w_{1,2}} +\Big(\frac{\lambda }{2}+ \frac{\omega }{2}\Big)\right)\nonumber\\
&&\times  \Bigg\{ \sum_{i_0=0}^{\beta _0} \frac{(-\beta _0)_{i_0}}{(1+\frac{\lambda }{2})_{i_0}(\gamma +\frac{\lambda }{2})_{i_0}}w_{1,2}^{i_0} \Bigg\} \tilde{\varepsilon }^2 \label{eq:104}\\
&& \mathrm{where}\hspace{.5cm} w_{1,2}=z\prod _{l=1}^{2} t_l u_l v_l \nonumber
\end{eqnarray}
By using similar process for the previous cases of integral forms of $y_1(x)$ and $y_2(x)$, the integral form of sub-power series expansion of $y_3(x)$ is
\begin{eqnarray}
 y_3(x)&=& c_0 x^{\lambda } \int_{0}^{1} dt_3\;t_3^{\frac{1}{2}+\frac{\lambda }{2}} \int_{0}^{1} du_3\;u_3^{\gamma -\frac{1}{2}+\frac{\lambda }{2}} \frac{1}{2\pi i}  \oint dv_3 \frac{\exp\left(-\frac{v_3}{(1-v_3)}z(1-t_3)(1-u_3)\right)}{v_3^{\beta _3+1}(1-v_3)} \left( w_{3,3}\partial _{w_{3,3}} +\Big( 1+ \frac{\lambda }{2}+ \frac{\omega }{2}\Big)\right)\nonumber\\
&&\times \int_{0}^{1} dt_2\;t_2^{\frac{\lambda }{2}} \int_{0}^{1} du_2\;u_2^{\gamma -1+\frac{\lambda }{2}} 
\frac{1}{2\pi i}  \oint dv_2 \frac{\exp\left(-\frac{v_2}{(1-v_2)}w_{3,3}(1-t_2)(1-u_2)\right)}{v_2^{\beta _2+1}(1-v_2)} 
\left( w_{2,3}\partial _{w_{2,3}} +\Big( \frac{1}{2}+ \frac{\lambda }{2}+ \frac{\omega }{2}\Big)\right)\nonumber\\
&&\times \int_{0}^{1} dt_1\;t_1^{-\frac{1}{2}+\frac{\lambda }{2}} \int_{0}^{1} du_1\;u_1^{\gamma -\frac{3}{2}+\frac{\lambda }{2}} 
\frac{1}{2\pi i}  \oint dv_1 \frac{\exp\left(-\frac{v_1}{(1-v_1)}w_{2,3}(1-t_1)(1-u_1)\right)}{v_1^{\beta _1+1}(1-v_1)}\left( w_{1,3}\partial _{w_{1,3}} +\Big( \frac{\lambda }{2}+ \frac{\omega }{2}\Big)\right)\nonumber\\
&&\times   \Bigg\{ \sum_{i_0=0}^{\beta _0 }\frac{(-\beta _0)_{i_0}}{(1+\frac{\lambda }{2})_{i_0}(\gamma +\frac{\lambda }{2})_{i_0}}  w_{1,3}^{i_0}\Bigg\} \tilde{\varepsilon }^3 \label{eq:105}
\end{eqnarray}
where
\begin{equation}
\begin{cases} \displaystyle{w_{3,3} =z\prod _{l=3}^{3} t_l u_l v_l} \cr
\displaystyle{w_{2,3} =z\prod _{l=2}^{3} t_l u_l v_l} \cr
 \displaystyle{w_{1,3}= z\prod _{l=1}^{3} t_l u_l v_l}
\end{cases}
\nonumber
\end{equation}
By repeating this process for all higher terms of integral forms of sub-summation $y_m(x)$ terms where $m \geq 4$, we obtain every integral forms of $y_m(x)$ terms. 
Since we substitute (\ref{eq:101a}), (\ref{eq:102}), (\ref{eq:104}), (\ref{eq:105}) and including all integral forms of $y_m(x)$ terms where $m \geq 4$ into (\ref{eq:100}), we obtain (\ref{eq:9}).
\qed
\end{pot} 
Let $\lambda = 0$ and $c_0= \frac{\Gamma( \beta _0+\gamma ) }{\Gamma (\gamma )}$ in (\ref{eq:9}). Apply (\ref{eq:4}) into the new (\ref{eq:9}).
\begin{rmk}
The integral representation of GCH equation of the first kind for polynomial which makes $B_n$ term terminated about $x=0$ as $\Omega = -2\mu (\beta _i+\frac{i}{2})$ where $i,\beta _i = 0,1,2,\cdots$ is
\begin{eqnarray}
 y(x)&=&  QW_{\beta _i}\left( \beta _i=-\frac{\Omega }{2\mu }-\frac{i}{2}, \gamma =\frac{1}{2}(1+\nu );\; \tilde{\varepsilon }= -\frac{1}{2}\varepsilon x;\; z=-\frac{1}{2}\mu x^2 \right)\nonumber\\
&=& F_{\beta _0}(\gamma ;z) + \sum_{n=1}^{\infty } \Bigg\{\prod _{j=0}^{n-1} \Bigg\{ \int_{0}^{1} dt_{n-j}\;t_{n-j}^{\frac{1}{2}(n-j)-1} \int_{0}^{1} du_{n-j}\;u_{n-j}^{\gamma +\frac{1}{2}(n-j)-2} \nonumber\\
&&\times \frac{1}{2\pi i}  \oint dv_{n-j} \frac{\exp\left(-\frac{v_{n-j}}{(1-v_{n-j})}w_{n-j+1,n}(1-t_{n-j})(1-u_{n-j})\right)}{v_{n-j}^{\beta _{n-j}+1}(1-v_{n-j})}\nonumber\\
&&\times \left( w_{n-j,n}\partial _{w_{n-j,n}} +\frac{1}{2}\Big(n-j-1+\omega \Big)\right) F_{\beta  _0}(\gamma ;w_{1,n})\Bigg\}\Bigg\}  \tilde{\varepsilon }^n  
\label{eq:11}
\end{eqnarray}
\end{rmk}
Confluent hypergeometric polynomial of the second kind is defined by
\begin{equation}
A_{\psi _0}(\gamma ;z)= \frac{\Gamma (\psi _0+2-\gamma )}{\Gamma (2-\gamma )} \sum_{n=0}^{\psi _0} \frac{(-\psi _0)_n}{n!\;(2-\gamma )_n}z^n
= \frac{\psi _0!}{2\pi i}\oint d v_0 \frac{\exp\left(-\frac{zv_0}{(1-v_0)}\right)}{v_0^{\psi _0+1}(1-v_0)^{2-\gamma }}
\label{eq:12}
\end{equation}
Put $c_0= \left( -\frac{1}{2}\mu \right)^{1-\gamma } \frac{\Gamma (\psi _0+2-\gamma )}{\Gamma (2-\gamma )}$ as $\lambda = 1-\nu = 2(1-\gamma )$ on (\ref{eq:9}) with replacing $\beta _i$ by $\psi _i$. apply (\ref{eq:12}) into the new (\ref{eq:9}).
\begin{rmk}
The integral representation of GCH equation of the second kind for polynomial which makes $B_n$ term terminated about $x=0$ as $\Omega = -2\mu (\psi _i +1-\gamma +\frac{i}{2})$ where $i, \psi _i = 0,1,2,\cdots$ is
\begin{eqnarray}
 y(x)&=&  RW_{\psi _i}\left( \psi _i=-\frac{\Omega }{2\mu }+\gamma -1-\frac{i}{2}, \gamma =\frac{1}{2}(1+\nu );\; \tilde{\varepsilon }= -\frac{1}{2}\varepsilon x;\; z=-\frac{1}{2}\mu x^2 \right) \nonumber\\
&=& z^{1-\gamma } \Bigg\{A_{\psi _0}(\gamma ;z) + \sum_{n=1}^{\infty } \Bigg\{\prod _{j=0}^{n-1} \Bigg\{ \int_{0}^{1} dt_{n-j}\;t_{n-j}^{\frac{1}{2}(n-j)-\gamma } \int_{0}^{1} du_{n-j}\;u_{n-j}^{\frac{1}{2}(n-j)-1} \nonumber\\
&&\times \frac{1}{2\pi i}  \oint dv_{n-j} \frac{\exp\left(-\frac{v_{n-j}}{(1-v_{n-j})}w_{n-j+1,n}(1-t_{n-j})(1-u_{n-j})\right)}{v_{n-j}^{\psi  _{n-j}+1}(1-v_{n-j})}\nonumber\\
&&\times \left( w_{n-j,n}\partial _{w_{n-j,n}} +\frac{1}{2}\Big(n-j+1-2\gamma +\omega \Big)\right) A_{\psi _0}(\gamma ;w_{1,n})\Bigg\}\Bigg\}  \tilde{\varepsilon }^n \Bigg\} 
\label{eq:13}
\end{eqnarray}
\end{rmk}

\subsection{Infinite series}
\begin{thm}
The general expression of the integral representation of GCH equation for infinite series is given by
\begin{eqnarray}
 y(x) &=& \sum_{n=0}^{\infty } y_n(x)= y_0(x)+ y_1(x)+ y_2(x)+ y_3(x)+ \cdots \nonumber\\
&=& c_0 x^{\lambda } \Bigg\{ \sum_{i_0=0}^{\infty} \frac{(\frac{\Omega }{2\mu }+\frac{\lambda }{2})_{i_0}}{(1+\frac{\lambda }{2})_{i_0}(\gamma +\frac{\lambda }{2})_{i_0}}z^{i_0}\nonumber\\
&& + \sum_{n=1}^{\infty } \Bigg\{\prod _{j=0}^{n-1} \Bigg\{ \int_{0}^{1} dt_{n-j}\;t_{n-j}^{\frac{1}{2}(n-j)-1+\frac{\lambda }{2}} \int_{0}^{1} du_{n-j}\;u_{n-j}^{\gamma +\frac{1}{2}(n-j)-2+\frac{\lambda }{2}} \nonumber\\
&&\times \frac{1}{2\pi i}  \oint dv_{n-j} \frac{\exp\left(-\frac{v_{n-j}}{(1-v_{n-j})}w_{n-j+1,n}(1-t_{n-j})(1-u_{n-j})\right)}{v_{n-j}^{-(\frac{\Omega }{2\mu }+\frac{n-j}{2}+ \frac{\lambda }{2})+1}(1-v_{n-j})}\nonumber\\
&&\times \left( w_{n-j,n}\partial _{w_{n-j,n}} +\frac{1}{2}\Big(n-j-1+\omega +\lambda \Big)\right) \Bigg\} \sum_{i_0=0}^{\infty} \frac{(\frac{\Omega }{2\mu }+\frac{\lambda }{2})_{i_0}}{(1+\frac{\lambda }{2})_{i_0}(\gamma +\frac{\lambda }{2})_{i_0}} w_{1,n}^{i_0}\Bigg\} \tilde{\varepsilon }^n \Bigg\}  \label{eq:14}
\end{eqnarray}
\end{thm}
\begin{pot} 
There is a generalized hypergeometric function which is written by
\begin{eqnarray}
L_j &=& \sum_{i_j= i_{j-1}}^{\infty } \frac{\left(\frac{\Omega }{2\mu }+\frac{j}{2}+\frac{\lambda }{2}\right)_{i_j}(1+\frac{j}{2}+\frac{\lambda }{2})_{i_{j-1}}(\frac{j}{2}+\gamma +\frac{\lambda }{2})_{i_{j-1}}}{\left(\frac{\Omega }{2\mu }+\frac{j}{2}+\frac{\lambda }{2}\right)_{i_{j-1}}(1+\frac{j}{2}+\frac{\lambda }{2})_{i_j}(\frac{j}{2}+\gamma +\frac{\lambda }{2})_{i_j}} z^{i_j}\nonumber\\
&=& z^{i_{j-1}} 
\sum_{l=0}^{\infty } \frac{B(i_{j-1}+\frac{j}{2}+\frac{\lambda }{2},l+1) B(i_{j-1}-1+\gamma +\frac{j}{2}+\frac{\lambda }{2},l+1)\left( \frac{\Omega }{2\mu }+\frac{j}{2}+\frac{\lambda }{2}+i_{j-1}\right)_l}{(i_{j-1}+\frac{j}{2}+\frac{\lambda }{2})^{-1}(i_{j-1}-1+\gamma +\frac{j}{2}+ \frac{\lambda }{2})^{-1}(1)_l \;l!} z^l \hspace{1cm}\label{eq:200}
\end{eqnarray}
Substitute (\ref{eq:2a}) and (\ref{eq:2b}) into (\ref{eq:200}), and divide $(i_{j-1}+\frac{j}{2}+\frac{\lambda }{2})(i_{j-1}-1+\gamma +\frac{j}{2}+ \frac{\lambda }{2})$ into the new (\ref{eq:200}).
\begin{eqnarray}
&& \frac{1}{(i_{j-1}+\frac{j}{2}+\frac{\lambda }{2})(i_{j-1}-1+\gamma +\frac{j}{2}+ \frac{\lambda }{2})}
 \sum_{i_j= i_{j-1}}^{\infty } \frac{\left(\frac{\Omega }{2\mu }+\frac{j}{2}+\frac{\lambda }{2}\right)_{i_j}(1+\frac{j}{2}+\frac{\lambda }{2})_{i_{j-1}}(\frac{j}{2}+\gamma +\frac{\lambda }{2})_{i_{j-1}}}{\left(\frac{\Omega }{2\mu }+\frac{j}{2}+\frac{\lambda }{2}\right)_{i_{j-1}}(1+\frac{j}{2}+\frac{\lambda }{2})_{i_j}(\frac{j}{2}+\gamma +\frac{\lambda }{2})_{i_j}} z^{i_j}\nonumber\\
&=&  \int_{0}^{1} dt_j\;t_j^{\frac{j}{2}-1+\frac{\lambda }{2}} \int_{0}^{1} du_j\;u_j^{\gamma -2+\frac{j}{2}+\frac{\lambda }{2}} (z t_j u_j)^{i_{j-1}}
 \sum_{l=0}^{\infty } \frac{\left( \frac{\Omega }{2\mu }+\frac{j}{2}+\frac{\lambda }{2}+i_{j-1}\right)_l}{(1)_l \;l!} [z(1-t_j)(1-u_j)]^l
 \label{eq:201}
\end{eqnarray}
Kummer function of the first kind is defined by
\begin{eqnarray}
M(a,b,z)&=& \sum_{n=0}^{\infty } \frac{(a)_n}{(b)_n n!} z^n = e^z M(b-a,b,-z) \nonumber\\
&=& -\frac{1}{2\pi i}\frac{\Gamma \left( 1-a\right)\Gamma \left(b\right)}{\Gamma \left(b-a\right) } \oint  dv_j\; e^{z v_j}(-v_j)^{a-1} \left( 1- v_j \right)^{b-a-1} \nonumber\\
&=&  \frac{\Gamma \left(a\right)}{2\pi i} \oint  dv_j\; e^{v_j}v_j^{-b} \left( 1-\frac{z}{v_j}\right)^{-a} \nonumber\\
&=& \frac{1}{2\pi i} \frac{\Gamma \left( 1-a\right)\Gamma \left(b\right)}{\Gamma \left(b-a\right) } \oint  dv_j\; e^{-\frac{z v_j}{1-v_j}}v_j^{a-1} \left( 1- v_j \right)^{-b} \label{eq:202}
\end{eqnarray}
Replace $a$, $b$ and $z$ by $\frac{\Omega }{2\mu }+\frac{j}{2}+\frac{\lambda }{2}+i_{j-1}$, 1 and $z(1-t_j)(1-u_j)$ in (\ref{eq:202}).
Take the new (\ref{eq:202}) into (\ref{eq:201}).
\begin{eqnarray}
Q_j &=& \frac{1}{(i_{j-1}+\frac{j}{2}+\frac{\lambda }{2})(i_{j-1}-1+\gamma +\frac{j}{2}+ \frac{\lambda }{2})}
 \sum_{i_j= i_{j-1}}^{\infty } \frac{\left(\frac{\Omega }{2\mu }+\frac{j}{2}+\frac{\lambda }{2}\right)_{i_j}(1+\frac{j}{2}+\frac{\lambda }{2})_{i_{j-1}}(\frac{j}{2}+\gamma +\frac{\lambda }{2})_{i_{j-1}}}{\left(\frac{\Omega }{2\mu }+\frac{j}{2}+\frac{\lambda }{2}\right)_{i_{j-1}}(1+\frac{j}{2}+\frac{\lambda }{2})_{i_j}(\frac{j}{2}+\gamma +\frac{\lambda }{2})_{i_j}} z^{i_j}\nonumber\\
&=& \int_{0}^{1} dt_j\;t_j^{\frac{j}{2}-1+\frac{\lambda }{2}} \int_{0}^{1} du_j\;u_j^{\gamma -2+\frac{j}{2}+\frac{\lambda }{2}} 
\frac{1}{2\pi i}  \oint dv_j \frac{\exp\left(-\frac{v_j}{(1-v_j)}z(1-t_j)(1-u_j)\right)}{v_j^{-\left( \frac{\Omega }{2\mu }+\frac{j}{2}+\frac{\lambda }{2}\right) +1}(1-v_j)}(z t_j u_j v_j)^{i_{j-1}} \hspace{1cm} \label{eq:203}  
\end{eqnarray}
In Ref.\cite{Chou2012i} the general expression of power series of GCH equation for infinite series is given by
\begin{eqnarray}
 y(x) &=& \sum_{n=0}^{\infty } y_{n}(x)= y_0(x)+ y_1(x)+ y_2(x)+ y_3(x)+\cdots \nonumber\\
&=& c_0 x^{\lambda } \Bigg\{\sum_{i_0=0}^{\infty } \frac{(\frac{\Omega }{2\mu }+ \frac{\lambda }{2})_{i_0}}{(1+\frac{\lambda }{2})_{i_0}(\gamma +\frac{\lambda }{2})_{i_0}}z^{i_0} 
+   \Bigg\{ \sum_{i_0=0}^{\infty }\frac{(i_0+\frac{\lambda }{2}+\frac{\omega }{2})}{(i_0+\frac{1}{2}+\frac{\lambda }{2})(i_0-\frac{1}{2}+\gamma +\frac{\lambda }{2})} \frac{(\frac{\Omega }{2\mu }+\frac{\lambda }{2})_{i_0}}{(1+\frac{\lambda }{2})_{i_0}(\gamma +\frac{\lambda }{2})_{i_0}}\nonumber\\
&&\times  \sum_{i_1=i_0}^{\infty }\frac{(\frac{\Omega }{2\mu }+\frac{1}{2}+\frac{\lambda }{2})_{i_1}(\frac{3}{2}+\frac{\lambda }{2})_{i_0}(\gamma +\frac{1}{2}+ \frac{\lambda }{2})_{i_0}}{(\frac{\Omega }{2\mu }+\frac{1}{2}+\frac{\lambda }{2})_{i_0}(\frac{3}{2}+\frac{\lambda }{2})_{i_1}(\gamma +\frac{1}{2}+\frac{\lambda }{2})_{i_1}} z^{i_1} \Bigg\} \tilde{\varepsilon }\nonumber\\
&&+ \sum_{n=2}^{\infty } \Bigg\{ \sum_{i_0=0}^{\infty } \frac{(i_0+\frac{\lambda }{2}+\frac{\omega }{2})}{(i_0+\frac{1}{2}+\frac{\lambda }{2})(i_0-\frac{1}{2}+\gamma +\frac{\lambda }{2})} \frac{(\frac{\Omega }{2\mu }+\frac{\lambda }{2})_{i_0}}{(1+\frac{\lambda }{2})_{i_0}(\gamma +\frac{\lambda }{2})_{i_0}} \nonumber\\
&&\times \prod _{k=1}^{n-1} \Bigg\{ \sum_{i_k=i_{k-1}}^{\infty } \frac{(i_k+\frac{\lambda }{2}+\frac{\omega }{2}+\frac{k}{2})}{(i_k+\frac{1}{2}+\frac{\lambda }{2}+\frac{k}{2})(i_k-\frac{1}{2}+\gamma + \frac{k}{2}+\frac{\lambda }{2})}  \frac{(\frac{\Omega }{2\mu }+\frac{k}{2}+\frac{\lambda }{2})_{i_k}(1+\frac{k}{2}+\frac{\lambda }{2})_{i_{k-1}}(\frac{k}{2}+\gamma +\frac{\lambda }{2})_{i_{k-1}}}{(\frac{\Omega }{2\mu }+\frac{k}{2}+\frac{\lambda }{2})_{i_{k-1}}(1+\frac{k}{2}+\frac{\lambda }{2})_{i_k}(\frac{k}{2}+\gamma +\frac{\lambda }{2})_{i_k}}\Bigg\} \nonumber\\
&&\times \sum_{i_n= i_{n-1}}^{\infty } \frac{(\frac{\Omega }{2\mu }+\frac{n}{2}+\frac{\lambda }{2})_{i_n}(1+\frac{n}{2}+\frac{\lambda }{2})_{i_{n-1}}(\frac{n}{2}+\gamma +\frac{\lambda }{2})_{i_{n-1}}}{(\frac{\Omega }{2\mu }+\frac{n}{2}+\frac{\lambda }{2})_{i_{n-1}}(1+\frac{n}{2}+\frac{\lambda }{2})_{i_n}(\frac{n}{2}+\gamma +\frac{\lambda }{2})_{i_n}} z^{i_n}\Bigg\} \tilde{\varepsilon }^n \Bigg\}
\label{eq:204}
\end{eqnarray}
In (\ref{eq:204}) sub-power series $y_0(x) $, $y_1(x)$, $y_2(x)$ and $y_3(x)$ of the GCH equation for infinite series  using 3TRF about $x=0$ are
\begin{subequations}
\begin{equation}
 y_0(x)= c_0 x^{\lambda } \sum_{i_0=0}^{\infty }\frac{\left(\frac{\Omega }{2\mu }+\frac{\lambda }{2}\right)_{i_0}}{(1+\frac{\lambda }{2})_{i_0}(\gamma +\frac{\lambda }{2})_{i_0}}  z^{i_0} \label{eq:205a}
\end{equation}
\begin{eqnarray}
 y_1(x)&=&  c_0 x^{\lambda } \Bigg\{ \sum_{i_0=0}^{\infty }\frac{(i_0+\frac{\lambda }{2}+\frac{\omega }{2})}{(i_0+\frac{1}{2}+\frac{\lambda }{2})(i_0-\frac{1}{2}+\gamma +\frac{\lambda }{2})} \frac{\left(\frac{\Omega }{2\mu }+\frac{\lambda }{2}\right)_{i_0}}{(1+\frac{\lambda }{2})_{i_0}(\gamma +\frac{\lambda }{2})_{i_0}}\nonumber\\
&&\times  \sum_{i_1=i_0}^{\infty } \frac{\left(\frac{\Omega }{2\mu }+\frac{1}{2}+\frac{\lambda }{2}\right)_{i_1}(\frac{3}{2}+\frac{\lambda }{2})_{i_0}(\gamma +\frac{1}{2}+ \frac{\lambda }{2})_{i_0}}{\left(\frac{\Omega }{2\mu }+\frac{1}{2}+\frac{\lambda }{2}\right)_{i_0}(\frac{3}{2}+\frac{\lambda }{2})_{i_1}(\gamma +\frac{1}{2}+\frac{\lambda }{2})_{i_1}} z^{i_1} \Bigg\}\tilde{\varepsilon } \label{eq:205b}
\end{eqnarray}
\begin{eqnarray}
 y_2(x) &=& c_0 x^{\lambda }\Bigg\{  \sum_{i_0=0}^{\infty }\frac{(i_0+\frac{\lambda }{2}+\frac{\omega }{2})}{(i_0+\frac{1}{2}+\frac{\lambda }{2})(i_0-\frac{1}{2}+\gamma +\frac{\lambda }{2})} \frac{\left(\frac{\Omega }{2\mu } +\frac{\lambda }{2}\right)_{i_0}}{(1+\frac{\lambda }{2})_{i_0}(\gamma +\frac{\lambda }{2})_{i_0}}\nonumber\\
&&\times  \sum_{i_1=i_0}^{\infty } \frac{(i_1+\frac{1}{2}+\frac{\lambda }{2}+\frac{\omega }{2})}{(i_1+1+\frac{\lambda }{2})(i_1+\gamma +\frac{\lambda }{2})} \frac{\left(\frac{\Omega }{2\mu }+\frac{1}{2}+\frac{\lambda }{2}\right)_{i_1}(\frac{3}{2}+\frac{\lambda }{2})_{i_0}(\gamma +\frac{1}{2}+ \frac{\lambda }{2})_{i_0}}{\left(\frac{\Omega }{2\mu }+\frac{1}{2}+\frac{\lambda }{2}\right)_{i_0}(\frac{3}{2}+\frac{\lambda }{2})_{i_1}(\gamma +\frac{1}{2}+\frac{\lambda }{2})_{i_1}} \nonumber\\
&&\times \sum_{i_2=i_1}^{\infty } \frac{\left(\frac{\Omega }{2\mu }+1+\frac{\lambda }{2}\right)_{i_2}(2+\frac{\lambda }{2})_{i_1}(\gamma +1+ \frac{\lambda }{2})_{i_1}}{\left(\frac{\Omega }{2\mu }+1+\frac{\lambda }{2}\right)_{i_1}(2+\frac{\lambda }{2})_{i_2}(\gamma +1+\frac{\lambda }{2})_{i_2}} z^{i_2} \Bigg\} \tilde{\varepsilon }^2 
\label{eq:205c}
\end{eqnarray}
\begin{eqnarray}
 y_3(x)&=&  c_0 x^{\lambda } \Bigg\{ \sum_{i_0=0}^{\infty }\frac{(i_0+\frac{\lambda }{2}+\frac{\omega }{2})}{(i_0+\frac{1}{2}+\frac{\lambda }{2})(i_0-\frac{1}{2}+\gamma +\frac{\lambda }{2})} \frac{\left(\frac{\Omega }{2\mu } +\frac{\lambda }{2}\right)_{i_0}}{(1+\frac{\lambda }{2})_{i_0}(\gamma +\frac{\lambda }{2})_{i_0}}\nonumber\\
&&\times  \sum_{i_1=i_0}^{\infty } \frac{(i_1+\frac{1}{2}+\frac{\lambda }{2}+\frac{\omega }{2})}{(i_1+1+\frac{\lambda }{2})(i_1+\gamma +\frac{\lambda }{2})} \frac{\left(\frac{\Omega }{2\mu }+\frac{1}{2}+\frac{\lambda }{2}\right)_{i_1}(\frac{3}{2}+\frac{\lambda }{2})_{i_0}(\gamma +\frac{1}{2}+ \frac{\lambda }{2})_{i_0}}{\left(\frac{\Omega }{2\mu }+\frac{1}{2}+\frac{\lambda }{2}\right)_{i_0}(\frac{3}{2}+\frac{\lambda }{2})_{i_1}(\gamma +\frac{1}{2}+\frac{\lambda }{2})_{i_1}} \nonumber\\
&&\times \sum_{i_2=i_1}^{\infty }\frac{(i_2+1+\frac{\lambda }{2}+\frac{\omega }{2})}{(i_2+\frac{3}{2}+\frac{\lambda }{2})(i_2+\frac{1}{2}+ \gamma +\frac{\lambda }{2})} \frac{\left(\frac{\Omega }{2\mu }+1+\frac{\lambda }{2}\right)_{i_2}(2+\frac{\lambda }{2})_{i_1}(\gamma +1+ \frac{\lambda }{2})_{i_1}}{\left(\frac{\Omega }{2\mu }+1+\frac{\lambda }{2}\right)_{i_1}(2+\frac{\lambda }{2})_{i_2}(\gamma +1+\frac{\lambda }{2})_{i_2}}\nonumber\\
&&\times \sum_{i_3=i_2}^{\infty } \frac{\left(\frac{\Omega }{2\mu }+\frac{3}{2}+\frac{\lambda }{2}\right)_{i_3}(\frac{5}{2}+\frac{\lambda }{2})_{i_2}(\gamma +\frac{3}{2}+ \frac{\lambda }{2})_{i_2}}{\left(\frac{\Omega }{2\mu }+\frac{3}{2}+\frac{\lambda }{2}\right)_{i_2}(\frac{5}{2}+\frac{\lambda }{2})_{i_3}(\gamma +\frac{3}{2}+\frac{\lambda }{2})_{i_3}} z^{i_3} \Bigg\} \tilde{\varepsilon }^3 
\label{eq:205d}
\end{eqnarray}
\end{subequations}
Put $j=1$ in (\ref{eq:203}). Take the new (\ref{eq:203}) into (\ref{eq:205b}).
\begin{eqnarray}
 y_1(x)&=& c_0 x^{\lambda }  \int_{0}^{1} dt_1\;t_1^{-\frac{1}{2}+\frac{\lambda }{2}} \int_{0}^{1} du_1\;u_1^{\gamma -\frac{3}{2}+\frac{\lambda }{2}} 
\frac{1}{2\pi i}  \oint dv_1 \frac{\exp\left(-\frac{v_1}{(1-v_1)}z(1-t_1)(1-u_1)\right)}{v_1^{-\left(\frac{\Omega }{2\mu }+\frac{1}{2}+\frac{\lambda }{2}\right) +1}(1-v_1)} \nonumber\\
&&\times \left\{ \sum_{i_0=0}^{\infty } \left(i_0+\frac{\omega }{2}+\frac{\lambda }{2}\right) \frac{\left(\frac{\Omega }{2\mu }+\frac{\lambda }{2}\right)_{i_0}}{(1+\frac{\lambda }{2})_{i_0}(\gamma +\frac{\lambda }{2})_{i_0}} (z t_1 u_1 v_1)^{i_0} \right\} \tilde{\varepsilon }\nonumber\\
&=& c_0 x^{\lambda }  \int_{0}^{1} dt_1\;t_1^{-\frac{1}{2}+\frac{\lambda }{2}} \int_{0}^{1} du_1\;u_1^{\gamma -\frac{3}{2}+\frac{\lambda }{2}} 
\frac{1}{2\pi i}  \oint dv_1 \frac{\exp\left(-\frac{v_1}{(1-v_1)}z(1-t_1)(1-u_1)\right)}{v_1^{-\left(\frac{\Omega }{2\mu }+\frac{1}{2}+\frac{\lambda }{2}\right)+1}(1-v_1)} \left( w_{1,1}\partial_{w_{1,1}} +\left( \frac{\omega }{2}+\frac{\lambda }{2}\right)\right)\nonumber\\
&&\times   \left\{ \sum_{i_0=0}^{\infty } \frac{\left(\frac{\Omega }{2\mu } +\frac{\lambda }{2}\right)_{i_0}}{(1+\frac{\lambda }{2})_{i_0}(\gamma +\frac{\lambda }{2})_{i_0}} w_{1,1}^{i_0} \right\} \tilde{\varepsilon }\label{eq:206}\\
&& \mathrm{where}\hspace{.5cm} w_{1,1}=z\prod _{l=1}^{1} t_l u_l v_l \nonumber
\end{eqnarray}
Put $j=2$ in (\ref{eq:203}). Take the new (\ref{eq:203}) into (\ref{eq:205c}). 
\begin{eqnarray}
 y_2(x)&=& c_0 x^{\lambda } \int_{0}^{1} dt_2\;t_2^{\frac{\lambda }{2}} \int_{0}^{1} du_2\;u_2^{\gamma -1+\frac{\lambda }{2}} 
\frac{1}{2\pi i}  \oint dv_2 \frac{\exp\left(-\frac{v_2}{(1-v_2)}z(1-t_2)(1-u_2)\right)}{v_2^{-\left(\frac{\Omega }{2\mu }+1+\frac{\lambda }{2}\right)+1}(1-v_2)}\left( w_{2,2}\partial _{w_{2,2}} +\Big( \frac{1}{2}+ \frac{\lambda }{2}+ \frac{\omega }{2}\Big)\right)\nonumber\\
&&\times  \Bigg\{ \sum_{i_0=0}^{\infty }\frac{(i_0+\frac{\lambda }{2}+\frac{\omega }{2})}{(i_0+\frac{1}{2}+\frac{\lambda }{2})(i_0-\frac{1}{2}+\gamma +\frac{\lambda }{2})} \frac{ \left(\frac{\Omega }{2\mu } +\frac{\lambda }{2}\right)_{i_0}}{(1+\frac{\lambda }{2})_{i_0}(\gamma +\frac{\lambda }{2})_{i_0}}\nonumber\\
&&\times  \sum_{i_1=i_0}^{\infty } \frac{ \left(\frac{\Omega }{2\mu }+1+\frac{\lambda }{2}\right)_{i_1}(\frac{3}{2}+\frac{\lambda }{2})_{i_0}(\gamma +\frac{1}{2}+ \frac{\lambda }{2})_{i_0}}{ \left(\frac{\Omega }{2\mu }+1+\frac{\lambda }{2}\right)_{i_0}(\frac{3}{2}+\frac{\lambda }{2})_{i_1}(\gamma +\frac{1}{2}+\frac{\lambda }{2})_{i_1}}w_{2,2}^{i_1} \Bigg\}\tilde{\varepsilon }^2 \label{eq:207}\\
&& \mathrm{where}\hspace{.5cm} w_{2,2}=z\prod _{l=2}^{2} t_l u_l v_l \nonumber
\end{eqnarray}
Put $j=1$ and $z =\overleftrightarrow {w}_{2,2}$ in (\ref{eq:203}). Take the new (\ref{eq:203}) into (\ref{eq:207}).
\begin{eqnarray}
 y_2(x)&=& c_0 x^{\lambda } \int_{0}^{1} dt_2\;t_2^{\frac{\lambda }{2}} \int_{0}^{1} du_2\;u_2^{\gamma -1+\frac{\lambda }{2}} 
\frac{1}{2\pi i}  \oint dv_2 \frac{\exp\left(-\frac{v_2}{(1-v_2)}z(1-t_2)(1-u_2)\right)}{v_2^{-\left(\frac{\Omega }{2\mu }+1+\frac{\lambda }{2}\right)+1}(1-v_2)} \left( w_{2,2}\partial _{w_{2,2}} +\Big( \frac{1}{2}+ \frac{\lambda }{2}+ \frac{\omega }{2}\Big)\right)\nonumber\\
&&\times  \int_{0}^{1} dt_1\;t_1^{-\frac{1}{2}+\frac{\lambda }{2}} \int_{0}^{1} du_1\;u_1^{\gamma -\frac{3}{2}+\frac{\lambda }{2}} 
\frac{1}{2\pi i}  \oint dv_1 \frac{\exp\left(-\frac{v_1}{(1-v_1)}w_{2,2}(1-t_1)(1-u_1)\right)}{v_1^{-\left(\frac{\Omega }{2\mu }+\frac{1}{2}+\frac{\lambda }{2}\right)+1}(1-v_1)} \left( w_{1,2}\partial _{w_{1,2}} +\Big(\frac{\lambda }{2}+ \frac{\omega }{2}\Big)\right)\nonumber\\
&&\times  \Bigg\{ \sum_{i_0=0}^{\infty } \frac{ \left(\frac{\Omega }{2\mu } +\frac{\lambda }{2}\right)_{i_0}}{(1+\frac{\lambda }{2})_{i_0}(\gamma +\frac{\lambda }{2})_{i_0}}w_{1,2}^{i_0} \Bigg\} \tilde{\varepsilon }^2 \label{eq:208}\\
&& \mathrm{where}\hspace{.5cm} w_{1,2}=z\prod _{l=1}^{2} t_l u_l v_l \nonumber
\end{eqnarray}
By using similar process for the previous cases of integral forms of $y_1(x)$ and $y_2(x)$, the integral form of sub-power series expansion of $y_3(x)$ is  
\begin{eqnarray}
 y_3(x)&=& c_0 x^{\lambda } \int_{0}^{1} dt_3\;t_3^{\frac{1}{2}+\frac{\lambda }{2}} \int_{0}^{1} du_3\;u_3^{\gamma -\frac{1}{2}+\frac{\lambda }{2}} \frac{1}{2\pi i}  \oint dv_3 \frac{\exp\left(-\frac{v_3}{(1-v_3)}z(1-t_3)(1-u_3)\right)}{v_3^{-\left(\frac{\Omega }{2\mu }+\frac{3}{2}+\frac{\lambda }{2}\right) +1}(1-v_3)} \left( w_{3,3}\partial _{w_{3,3}} +\Big( 1+ \frac{\lambda }{2}+ \frac{\omega }{2}\Big)\right)\nonumber\\
&&\times \int_{0}^{1} dt_2\;t_2^{\frac{\lambda }{2}} \int_{0}^{1} du_2\;u_2^{\gamma -1+\frac{\lambda }{2}} 
\frac{1}{2\pi i}  \oint dv_2 \frac{\exp\left(-\frac{v_2}{(1-v_2)}w_{3,3}(1-t_2)(1-u_2)\right)}{v_2^{-\left(\frac{\Omega }{2\mu }+1+\frac{\lambda }{2}\right) +1}(1-v_2)} 
\left( w_{2,3}\partial _{w_{2,3}} +\Big( \frac{1}{2}+ \frac{\lambda }{2}+ \frac{\omega }{2}\Big)\right)\nonumber\\
&&\times \int_{0}^{1} dt_1\;t_1^{-\frac{1}{2}+\frac{\lambda }{2}} \int_{0}^{1} du_1\;u_1^{\gamma -\frac{3}{2}+\frac{\lambda }{2}} 
\frac{1}{2\pi i}  \oint dv_1 \frac{\exp\left(-\frac{v_1}{(1-v_1)}w_{2,3}(1-t_1)(1-u_1)\right)}{v_1^{-\left(\frac{\Omega }{2\mu }+\frac{1}{2}+\frac{\lambda }{2}\right)+1}(1-v_1)}\left( w_{1,3}\partial _{w_{1,3}} +\Big( \frac{\lambda }{2}+ \frac{\omega }{2}\Big)\right)\nonumber\\
&&\times   \Bigg\{ \sum_{i_0=0}^{\infty }\frac{ \left(\frac{\Omega }{2\mu } +\frac{\lambda }{2}\right)_{i_0}}{(1+\frac{\lambda }{2})_{i_0}(\gamma +\frac{\lambda }{2})_{i_0}}  w_{1,3}^{i_0}\Bigg\} \tilde{\varepsilon }^3 \label{eq:209}
\end{eqnarray}
where
\begin{equation}
\begin{cases} \displaystyle{w_{3,3} =z\prod _{l=3}^{3} t_l u_l v_l} \cr
\displaystyle{w_{2,3} =z\prod _{l=2}^{3} t_l u_l v_l} \cr
 \displaystyle{w_{1,3}= z\prod _{l=1}^{3} t_l u_l v_l}
\end{cases}
\nonumber
\end{equation}
By repeating this process for all higher terms of integral forms of sub-summation $y_m(x)$ terms where $m \geq 4$, we obtain every integral forms of $y_m(x)$ terms. 
Since we substitute (\ref{eq:205a}), (\ref{eq:206}), (\ref{eq:208}), (\ref{eq:209}) and including all integral forms of $y_m(x)$ terms where $m \geq 4$ into (\ref{eq:204}), we obtain (\ref{eq:14}).
\qed
\end{pot} 
Let $\lambda = 0$ and $c_0= \frac{\Gamma( \gamma - \frac{\Omega }{2\mu }) }{\Gamma (\gamma )}$ in (\ref{eq:14}). And apply (\ref{eq:202}) into the new (\ref{eq:14}).
\begin{rmk}
The integral representation of GCH equation of the first kind for infinite series about $x=0$ for infinite series is
\begin{eqnarray}
 y(x) &=&  QW \left( \gamma =\frac{1}{2}(1+\nu );\; \tilde{\varepsilon }= -\frac{1}{2}\varepsilon x;\; z=-\frac{1}{2}\mu x^2 \right)\nonumber\\
&=& \frac{\Gamma( \gamma - \frac{\Omega }{2\mu }) }{\Gamma (\gamma )} \Bigg\{  M\left( \frac{\Omega }{2\mu },\gamma ,z\right)
+ \sum_{n=1}^{\infty } \Bigg\{\prod _{j=0}^{n-1} \Bigg\{ \int_{0}^{1} dt_{n-j}\;t_{n-j}^{\frac{1}{2}(n-j)-1} \int_{0}^{1} du_{n-j}\;u_{n-j}^{\gamma +\frac{1}{2}(n-j)-2} \nonumber\\
&&\times \frac{1}{2\pi i}  \oint dv_{n-j} \frac{\exp\left(-\frac{v_{n-j}}{(1-v_{n-j})}w_{n-j+1,n}(1-t_{n-j})(1-u_{n-j})\right)}{v_{n-j}^{-\frac{\Omega }{2\mu } -\frac{1}{2}(n-j)+1}(1-v_{n-j})}\left( w_{n-j,n}\partial _{w_{n-j,n}} +\frac{1}{2}\Big(n-j-1+\omega \Big)\right) \Bigg\}\nonumber\\
&&\times  M\left( \frac{\Omega }{2\mu },\gamma ,w_{1,n}\right) \Bigg\}  \tilde{\varepsilon }^n \Bigg\} 
\label{eq:16}
\end{eqnarray}
\end{rmk}
Put $c_0= \left( -\frac{1}{2}\mu \right)^{1-\gamma }\frac{\Gamma (1-\frac{\Omega }{2\mu })}{\Gamma (2-\gamma )}$ as $\lambda = 1-\nu = 2(1-\gamma )$ on (\ref{eq:14}). And apply (\ref{eq:202}) into the new (\ref{eq:14}).
\begin{rmk}
The integral representation of GCH equation of the second kind for infinite series about $x=0$ for infinite series is
\begin{eqnarray}
 y(x)&=& RW\left(\gamma =\frac{1}{2}(1+\nu );\; \tilde{\varepsilon }= -\frac{1}{2}\varepsilon x;\; z=-\frac{1}{2}\mu x^2 \right) \nonumber\\
&=& z^{1-\gamma }\frac{\Gamma (1-\frac{\Omega }{2\mu })}{\Gamma (2-\gamma )} \Bigg\{M\left( \frac{\Omega }{2\mu }+1-\gamma ,2-\gamma ,z\right)
+ \sum_{n=1}^{\infty } \Bigg\{\prod _{j=0}^{n-1} \Bigg\{ \int_{0}^{1} dt_{n-j}\;t_{n-j}^{\frac{1}{2}(n-j)-\gamma } \nonumber\\
&&\times \int_{0}^{1} du_{n-j}\;u_{n-j}^{\frac{1}{2}(n-j)-1} \frac{1}{2\pi i}  \oint dv_{n-j} \frac{\exp\left(-\frac{v_{n-j}}{(1-v_{n-j})}w_{n-j+1,n}(1-t_{n-j})(1-u_{n-j})\right)}{v_{n-j}^{-\frac{\Omega }{2\mu } +\gamma -\frac{1}{2}(n-j)}(1-v_{n-j})}\nonumber\\
&&\times \left( w_{n-j,n}\partial _{w_{n-j,n}} +\frac{1}{2}\Big(n-j+1-2\gamma +\omega \Big)\right)\Bigg\} M\left( \frac{\Omega }{2\mu }+1-\gamma ,2-\gamma ,w_{1,n}\right) \Bigg\} \tilde{\varepsilon }^n \Bigg\}  \label{eq:17}
\end{eqnarray}
\end{rmk}
\section{Generating function for the polynomial which makes $B_n$ term terminated}
Now let's investigate the generating function for the GCH polynomials of the first and second kinds. 
\begin{defn}
I define that
\begin{equation}
\begin{cases} \displaystyle { s_{a,b}} = \begin{cases} \displaystyle {  s_a\cdot s_{a+1}\cdot s_{a+2}\cdots s_{b-2}\cdot s_{b-1}\cdot s_b}\;\;\mbox{where}\;a>b \cr
s_a \;\;\mbox{only}\;\mbox{if}\;a=b\end{cases}
\cr
\displaystyle {w_{a,b}^{\ast } = z s_{a,\infty } \prod _{l=a}^{b} t_l u_l} \;\;\mbox{where}\;a\leq b 
\end{cases}
\label{eq:36}
\end{equation}
where
\begin{equation}
a,b\in \mathbb{N}_{0} \nonumber
\end{equation}
\end{defn}
And I have
\begin{equation}
\sum_{\beta _i=\beta _j}^{\infty } s_i^{\beta _i} = \frac{s_i^{\beta _j}}{(1-s_i)}
\label{eq:37}
\end{equation}
Acting the summation operator $\displaystyle{ \sum_{\beta _0=0}^{\infty } \frac{s_0^{\beta _0}}{\beta _0!}\prod _{n=1}^{\infty } \left\{ \sum_{\beta _n=\beta _{n-1}}^{\infty } s_n^{\beta _n}\right\}}$ on (\ref{eq:9}) where $|s_i|<1$ as $i=0,1,2,\cdots$ by using (\ref{eq:36}) and (\ref{eq:37}). 
\begin{thm}
The general expression of the generating function for the GCH polynomial which makes $B_n$ term terminated is given by
\begin{eqnarray}
&&\sum_{\beta _0=0}^{\infty } \frac{s_0^{\beta _0}}{\beta _0!}\prod _{n=1}^{\infty } \left\{ \sum_{\beta _n=\beta _{n-1}}^{\infty } s_n^{\beta _n}\right\}y(x) \nonumber\\
&&= \prod_{k=1}^{\infty } \frac{1}{(1-s_{k,\infty })} \mathbf{\Upsilon}(\lambda; s_{0,\infty } ;z)  \nonumber\\
&&+ \left\{ \prod_{k=1}^{\infty } \frac{1}{(1-s_{k,\infty })} \int_{0}^{1} dt_1\;t_1^{-\frac{1}{2}+\frac{\lambda }{2}} \int_{0}^{1} du_1\;u_1^{\gamma -\frac{3}{2}+\frac{\lambda }{2}}\right. \nonumber\\
&&\times \left. \exp\left(-\frac{s_{1,\infty }}{(1-s_{1,\infty })}z(1-t_1)(1-u_1)\right)  \left( w_{1,1}^{\ast }\partial_{w_{1,1}^{\ast }} +\left( \frac{\omega }{2}+\frac{\lambda }{2}\right)\right) \mathbf{\Upsilon}(\lambda ; s_0;w_{1,1}^{\ast })\right\} \tilde{\varepsilon } \nonumber\\
&&+ \sum_{n=2}^{\infty } \Bigg\{ \prod_{k=n}^{\infty } \frac{1}{(1-s_{k,\infty })} \int_{0}^{1} dt_n\;t_n^{\frac{n}{2}-1+\frac{\lambda }{2}} \int_{0}^{1} du_n \;u_n^{\gamma -2+\frac{n}{2}+\frac{\lambda }{2}} \exp\left(-\frac{s_{n,\infty }}{(1-s_{n,\infty })}z(1-t_n)(1-u_n)\right)\nonumber\\
&&\times \Bigg( w_{n,n}^{\ast }\partial _{w_{n,n}^{\ast }}+ \Big(\frac{1}{2}(n-1)+\frac{\omega }{2}+ \frac{\lambda }{2} \Big) \Bigg) \nonumber\\
&&\times  \prod_{j=1}^{n-1} \Bigg\{ \int_{0}^{1} dt_{n-j}\;t_{n-j}^{\frac{1}{2}(n-j)-1+\frac{\lambda }{2}} \int_{0}^{1} du_{n-j} \;u_{n-j}^{\gamma -2+\frac{1}{2}(n-j)+\frac{\lambda }{2}}\frac{\exp\left(-\frac{s_{n-j}}{(1-s_{n-j})}w_{n-j+1,n}^{\ast }(1-t_{n-j})(1-u_{n-j})\right)}{(1-s_{n-j})} \nonumber\\
&&\times \Bigg( w_{n-j,n}^{\ast }\partial _{w_{n-j,n}^{\ast }}+ \Big(\frac{1}{2}(n-j-1)+\frac{\omega }{2}+ \frac{\lambda }{2} \Big)\Bigg)\Bigg\}
 \mathbf{\Upsilon}(\lambda; s_0 ;w_{1,n}^{\ast }) \Bigg\} \tilde{\varepsilon }^n  \label{eq:38}
\end{eqnarray}
where
\begin{equation}
\begin{cases} 
{ \displaystyle \mathbf{\Upsilon}(\lambda; s_{0,\infty } ;z)= \sum_{\beta _0=0}^{\infty } \frac{s_{0,\infty }^{\beta _0}}{\beta _0!} \left\{  c_0 x^{\lambda }  \sum_{i_0=0}^{\beta _0} \frac{(-\beta _0)_{i_0}}{(1+\frac{\lambda }{2})_{i_0}(\gamma +\frac{\lambda }{2})_{i_0}}z^{i_0} \right\} }
\cr
{ \displaystyle \mathbf{\Upsilon}(\lambda ; s_0;w_{1,1}^{\ast }) =  \sum_{\beta _0=0}^{\infty }\frac{s_0^{\beta _0}}{\beta _0!}\left\{  c_0 x^{\lambda } \sum_{i_0=0}^{\beta _0} \frac{(-\beta _0)_{i_0}}{(1+\frac{\lambda }{2})_{i_0}(\gamma +\frac{\lambda }{2})_{i_0}} (w_{1,1}^{\ast })^{i_0}\right\} } \cr
{ \displaystyle \mathbf{\Upsilon}(\lambda; s_0 ;w_{1,n}^{\ast }) = \sum_{\beta _0=0}^{\infty }\frac{s_0^{\beta _0}}{\beta _0!} \left\{  c_0 x^{\lambda } \sum_{i_0=0}^{\beta _0}\frac{(-\beta _0)_{i_0}}{(1+\frac{\lambda }{2})_{i_0}(\gamma +\frac{\lambda }{2})_{i_0}}  (w_{1,n}^{\ast })^{i_0}\right\}}
\end{cases}\nonumber 
\end{equation}
\end{thm}
\begin{pot} 
Acting the summation operator $\displaystyle{ \sum_{\beta _0=0}^{\infty } \frac{s_0^{\beta _0}}{\beta _0!}\prod _{n=1}^{\infty } \left\{ \sum_{\beta _n=\beta _{n-1}}^{\infty } s_n^{\beta _n}\right\}}$ on the general expression of the integral representation of the GCH polynomial which makes $B_n$ term terminated $y(x)$,
\begin{eqnarray}
&& \sum_{\beta _0=0}^{\infty } \frac{s_0^{\beta _0}}{\beta _0!}\prod _{n=1}^{\infty } \left\{ \sum_{\beta _n=\beta _{n-1}}^{\infty } s_n^{\beta _n}\right\} y(x) \nonumber\\
&&=  \sum_{\beta _0=0}^{\infty } \frac{s_0^{\beta _0}}{\beta _0!}\prod _{n=1}^{\infty } \left\{ \sum_{\beta _n=\beta _{n-1}}^{\infty } s_n^{\beta _n}\right\} \left\{ y_0(x)+y_1(x)+y_2(x)+y_3(x)+ \cdots \right\} \label{eq:106.1}
\end{eqnarray}
Acting the summation operator $\displaystyle{ \sum_{\beta _0=0}^{\infty } \frac{s_0^{\beta _0}}{\beta _0!}\prod _{n=1}^{\infty } \left\{ \sum_{\beta _n=\beta _{n-1}}^{\infty } s_n^{\beta _n}\right\}}$ on (\ref{eq:101a}),
\begin{eqnarray}
&&\sum_{\beta _0=0}^{\infty } \frac{s_0^{\beta _0}}{\beta _0!}\prod _{n=1}^{\infty } \left\{ \sum_{\beta _n=\beta _{n-1}}^{\infty } s_n^{\beta _n}\right\}y_0(x)\nonumber\\
&& = \prod_{k=1}^{\infty } \frac{1}{(1-s_{k,\infty })} \sum_{\beta _0=0}^{\infty } \frac{s_{0,\infty }^{\beta _0}}{\beta _0!}\Bigg\{ c_0 x^{\lambda } \sum_{i_0=0}^{\beta _0} \frac{(-\beta _0)_{i_0}}{(1+\frac{\lambda }{2})_{i_0}(\gamma +\frac{\lambda }{2})_{i_0}}z^{i_0} \Bigg\} \label{eq:106}
\end{eqnarray}
Acting the summation operator $\displaystyle{ \sum_{\beta _0=0}^{\infty } \frac{s_0^{\beta _0}}{\beta _0!}\prod _{n=1}^{\infty } \left\{ \sum_{\beta _n=\beta _{n-1}}^{\infty } s_n^{\beta _n}\right\}}$ on (\ref{eq:102}),
\begin{eqnarray}
&&\sum_{\beta _0=0}^{\infty } \frac{s_0^{\beta _0}}{\beta _0!}\prod _{n=1}^{\infty } \left\{ \sum_{\beta _n=\beta _{n-1}}^{\infty } s_n^{\beta _n}\right\}y_1(x) \nonumber\\
&&=  \prod_{k=2}^{\infty } \frac{1}{(1-s_{k,\infty })} \int_{0}^{1} dt_1\;t_1^{-\frac{1}{2}+\frac{\lambda }{2}} \int_{0}^{1} du_1\;u_1^{\gamma -\frac{3}{2}+\frac{\lambda }{2}} 
  \frac{1}{2\pi i}  \oint dv_1 \frac{\exp\left(-\frac{v_1}{(1-v_1)}z(1-t_1)(1-u_1)\right)}{v_1(1-v_1)}\sum_{\beta _1=\beta _0}^{\infty }\left( \frac{s_{1,\infty }}{v_1}\right)^{\beta _1}\nonumber\\
&&\times  \left( w_{1,1}\partial_{w_{1,1}} +\left( \frac{\omega }{2}+\frac{\lambda }{2}\right)\right)\sum_{\beta _0=0}^{\infty }\frac{s_0^{\beta _0}}{\beta _0!} \Bigg\{ c_0 x^{\lambda} \sum_{i_0=0}^{\beta _0} \frac{(-\beta _0)_{i_0}}{(1+\frac{\lambda }{2})_{i_0}(\gamma +\frac{\lambda }{2})_{i_0}} w_{1,1}^{i_0}\Bigg\} \tilde{\varepsilon } \label{eq:107}
\end{eqnarray}
Replace $\beta _i$, $\beta _j$ and $s_i$ by $\beta _1$, $\beta _0$ and ${ \displaystyle \frac{s_{1,\infty }}{v_1}}$ in (\ref{eq:37}). Take the new (\ref{eq:37}) into (\ref{eq:107}).
\begin{eqnarray}
&&\sum_{\beta _0=0}^{\infty } \frac{s_0^{\beta _0}}{\beta _0!}\prod _{n=1}^{\infty } \left\{ \sum_{\beta _n=\beta _{n-1}}^{\infty } s_n^{\beta _n}\right\} y_1(x) \nonumber\\
&&= \prod_{k=2}^{\infty } \frac{1}{(1-s_{k,\infty })} \int_{0}^{1} dt_1\;t_1^{-\frac{1}{2}+\frac{\lambda }{2}} \int_{0}^{1} du_1\;u_1^{\gamma -\frac{3}{2}+\frac{\lambda }{2}} 
  \frac{1}{2\pi i}  \oint dv_1 \frac{\exp\left(-\frac{v_1}{(1-v_1)}z(1-t_1)(1-u_1)\right)}{(1-v_1)(v_1-s_{1,\infty})}\nonumber\\
&&\times  \left( w_{1,1}\partial_{w_{1,1}} +\left( \frac{\omega }{2}+\frac{\lambda }{2}\right)\right)\sum_{\beta _0=0}^{\infty }\frac{1}{\beta _0!}\left(\frac{s_{0,\infty }}{v_1}\right)^{\beta _0} \Bigg\{ c_0 x^{\lambda} \sum_{i_0=0}^{\beta _0} \frac{(-\beta _0)_{i_0}}{(1+\frac{\lambda }{2})_{i_0}(\gamma +\frac{\lambda }{2})_{i_0}} w_{1,1}^{i_0} \Bigg\}\tilde{\varepsilon }  \label{eq:108}
\end{eqnarray}
By using Cauchy's integral formula, the contour integrand has poles at $v_1 =1$ or $s_{1,\infty}$,
and $s_{1,\infty}$ is only inside the unit circle. As we compute the residue there in (\ref{eq:108}) we obtain
\begin{eqnarray}
&&\sum_{\beta _0=0}^{\infty } \frac{s_0^{\beta _0}}{\beta _0!}\prod _{n=1}^{\infty } \left\{ \sum_{\beta _n=\beta _{n-1}}^{\infty } s_n^{\beta _n}\right\} y_1(x) \nonumber\\
&&=  \prod_{k=1}^{\infty } \frac{1}{(1-s_{k,\infty })} \int_{0}^{1} dt_1\;t_1^{-\frac{1}{2}+\frac{\lambda }{2}} \int_{0}^{1} du_1\;u_1^{\gamma -\frac{3}{2}+\frac{\lambda }{2}} 
  \exp\left(-\frac{s_{1,\infty }}{(1-s_{1,\infty })}z(1-t_1)(1-u_1)\right)\nonumber\\
&&\times  \left( w_{1,1}^{\ast }\partial_{w_{1,1}^{\ast }} +\left( \frac{\omega }{2}+\frac{\lambda }{2}\right)\right)\sum_{\beta _0=0}^{\infty }\frac{s_0^{\beta _0}}{\beta _0!}\Bigg\{ c_0 x^{\lambda}\sum_{i_0=0}^{\beta _0} \frac{(-\beta _0)_{i_0}}{(1+\frac{\lambda }{2})_{i_0}(\gamma +\frac{\lambda }{2})_{i_0}} (w_{1,1}^{\ast })^{i_0} \Bigg\}\tilde{\varepsilon }  \label{eq:109}
\end{eqnarray}
where
\begin{eqnarray}
w_{1,1}^{\ast } &=& z s_{1,\infty } \prod _{l=1}^{1} t_l u_l\nonumber
\end{eqnarray}
Acting the summation operator $\displaystyle{ \sum_{\beta _0=0}^{\infty } \frac{s_0^{\beta _0}}{\beta _0!}\prod _{n=1}^{\infty } \left\{ \sum_{\beta _n=\beta _{n-1}}^{\infty } s_n^{\beta _n}\right\}}$ on (\ref{eq:104}),
\begin{eqnarray}
&&\sum_{\beta _0=0}^{\infty } \frac{s_0^{\beta _0}}{\beta _0!}\prod _{n=1}^{\infty } \left\{ \sum_{\beta _n=\beta _{n-1}}^{\infty } s_n^{\beta _n}\right\}y_2(x) \nonumber\\
&&=  \prod_{k=3}^{\infty } \frac{1}{(1-s_{k,\infty })} \int_{0}^{1} dt_2\;t_2^{\frac{\lambda }{2}} \int_{0}^{1} du_2\;u_2^{\gamma -1+\frac{\lambda }{2}} \nonumber\\
&&\times \frac{1}{2\pi i}  \oint dv_2 \frac{\exp\left(-\frac{v_2}{(1-v_2)}z(1-t_2)(1-u_2)\right)}{v_2(1-v_2)}\sum_{\beta _2=\beta _1}^{\infty }\left( \frac{s_{2,\infty }}{v_2}\right)^{\beta _2}\left( w_{2,2}\partial _{w_{2,2}} +\Big( \frac{1}{2}+ \frac{\lambda }{2}+ \frac{\omega }{2}\Big)\right)\nonumber\\
&&\times  \int_{0}^{1} dt_1\;t_1^{-\frac{1}{2}+\frac{\lambda }{2}} \int_{0}^{1} du_1\;u_1^{\gamma -\frac{3}{2}+\frac{\lambda }{2}} 
\frac{1}{2\pi i}  \oint dv_1 \frac{\exp\left(-\frac{v_1}{(1-v_1)}w_{2,2}(1-t_1)(1-u_1)\right)}{v_1 (1-v_1)} \nonumber\\
&&\times \sum_{\beta _1=\beta _0}^{\infty } \left( \frac{s_1}{v_1}\right)^{\beta _1}\left( w_{1,2}\partial _{w_{1,2}} +\Big(\frac{\lambda }{2}+ \frac{\omega }{2}\Big)\right)
 \sum_{\beta _0=0}^{\infty }\frac{s_0^{\beta _0}}{\beta _0!}\Bigg\{c_0 x^{\lambda} \sum_{i_0=0}^{\beta _0} \frac{(-\beta _0)_{i_0}}{(1+\frac{\lambda }{2})_{i_0}(\gamma +\frac{\lambda }{2})_{i_0}}w_{1,2}^{i_0} \Bigg\} \tilde{\varepsilon }^2\label{eq:110}
\end{eqnarray}
Replace $\beta _i$, $\beta _j$ and $s_i$ by $\beta _2$, $\beta _1$ and ${ \displaystyle \frac{s_{2,\infty }}{v_2}}$ in (\ref{eq:37}). Take the new (\ref{eq:37}) into (\ref{eq:110}).
\begin{eqnarray}
&&\sum_{\beta _0=0}^{\infty } \frac{s_0^{\beta _0}}{\beta _0!}\prod _{n=1}^{\infty } \left\{ \sum_{\beta _n=\beta _{n-1}}^{\infty } s_n^{\beta _n}\right\}y_2(x) \nonumber\\
&&= \prod_{k=3}^{\infty } \frac{1}{(1-s_{k,\infty })} \int_{0}^{1} dt_2\;t_2^{\frac{\lambda }{2}} \int_{0}^{1} du_2\;u_2^{\gamma -1+\frac{\lambda }{2}} \nonumber\\
&&\times \frac{1}{2\pi i}  \oint dv_2 \frac{\exp\left(-\frac{v_2}{(1-v_2)}z(1-t_2)(1-u_2)\right)}{(1-v_2)(v_2-s_{2,\infty })}\left( w_{2,2}\partial _{w_{2,2}} +\Big( \frac{1}{2}+ \frac{\lambda }{2}+ \frac{\omega }{2}\Big)\right)\nonumber\\
&&\times  \int_{0}^{1} dt_1\;t_1^{-\frac{1}{2}+\frac{\lambda }{2}} \int_{0}^{1} du_1\;u_1^{\gamma -\frac{3}{2}+\frac{\lambda }{2}} 
\frac{1}{2\pi i}  \oint dv_1 \frac{\exp\left(-\frac{v_1}{(1-v_1)}w_{2,2}(1-t_1)(1-u_1)\right)}{v_1 (1-v_1)} \nonumber \\
&&\times \sum_{\beta _1=\beta _0}^{\infty } \left( \frac{s_{1,\infty }}{v_1 v_2}\right)^{\beta _1}\left( w_{1,2}\partial _{w_{1,2}} +\Big(\frac{\lambda }{2}+ \frac{\omega }{2}\Big)\right)
\sum_{\beta _0=0}^{\infty }\frac{s_0^{\beta _0}}{\beta _0!} \Bigg\{c_0 x^{\lambda} \sum_{i_0=0}^{\beta _0} \frac{(-\beta _0)_{i_0}}{(1+\frac{\lambda }{2})_{i_0}(\gamma +\frac{\lambda }{2})_{i_0}}w_{1,2}^{i_0} \Bigg\}\tilde{\varepsilon }^2
\label{eq:111}
\end{eqnarray}
By using Cauchy's integral formula, the contour integrand has poles at $v_2 =1$ or $s_{2,\infty}$,
and $s_{2,\infty}$ is only inside the unit circle. As we compute the residue there in (\ref{eq:111}) we obtain
\begin{eqnarray}
&&\sum_{\beta _0=0}^{\infty } \frac{s_0^{\beta _0}}{\beta _0!}\prod _{n=1}^{\infty } \left\{ \sum_{\beta _n=\beta _{n-1}}^{\infty } s_n^{\beta _n}\right\}y_2(x) \nonumber\\
&&=   \prod_{k=2}^{\infty } \frac{1}{(1-s_{k,\infty })} \int_{0}^{1} dt_2\;t_2^{\frac{\lambda }{2}} \int_{0}^{1} du_2\;u_2^{\gamma -1+\frac{\lambda }{2}} \nonumber\\
&&\times  \exp\left(-\frac{s_{2,\infty }}{(1-s_{2,\infty })}z(1-t_2)(1-u_2)\right)\left( w_{2,2}^{\ast }\partial _{w_{2,2}^{\ast }} +\Big( \frac{1}{2}+ \frac{\lambda }{2}+ \frac{\omega }{2}\Big)\right)\nonumber\\
&&\times  \int_{0}^{1} dt_1\;t_1^{-\frac{1}{2}+\frac{\lambda }{2}} \int_{0}^{1} du_1\;u_1^{\gamma -\frac{3}{2}+\frac{\lambda }{2}} 
\frac{1}{2\pi i}  \oint dv_1 \frac{\exp\left(-\frac{v_1}{(1-v_1)}w_{2,2}^{\ast }(1-t_1)(1-u_1)\right)}{v_1(1-v_1)} \sum_{\beta _1=\beta _0}^{\infty } \left( \frac{s_1}{v_1}\right)^{\beta _1}\nonumber \\
&&\times \left( \ddot{w}_{1,2}\partial _{\ddot{w}_{1,2}} +\Big(\frac{\lambda }{2}+ \frac{\omega }{2}\Big)\right)
 \sum_{\beta _0=0}^{\infty }\frac{s_0^{\beta _0}}{\beta _0!} \Bigg\{ c_0 x^{\lambda}\sum_{i_0=0}^{\beta _0} \frac{(-\beta _0)_{i_0}}{(1+\frac{\lambda }{2})_{i_0}(\gamma +\frac{\lambda }{2})_{i_0}}\ddot{w}_{1,2}^{i_0}\Bigg\}\tilde{\varepsilon }^2 \label{eq:112}
\end{eqnarray}
where
\begin{eqnarray}
w_{2,2}^{\ast } &=& z s_{2,\infty } \prod _{l=2}^{2} t_l u_l\hspace{2cm}\ddot{w}_{1,2} = z s_{2,\infty } v_1\prod _{l=1}^{2} t_l u_l\nonumber
\end{eqnarray}
Replace $\beta _i$, $\beta _j$ and $s_i$ by $\beta _1$, $\beta _0$ and ${ \displaystyle \frac{s_1}{v_1}}$ in (\ref{eq:37}). Take the new (\ref{eq:37}) into (\ref{eq:112}).
\begin{eqnarray}
&&\sum_{\beta _0=0}^{\infty } \frac{s_0^{\beta _0}}{\beta _0!}\prod _{n=1}^{\infty } \left\{ \sum_{\beta _n=\beta _{n-1}}^{\infty } s_n^{\beta _n}\right\}y_2(x) \nonumber\\
&&= \prod_{k=2}^{\infty } \frac{1}{(1-s_{k,\infty })} \int_{0}^{1} dt_2\;t_2^{\frac{\lambda }{2}} \int_{0}^{1} du_2\;u_2^{\gamma -1+\frac{\lambda }{2}} \nonumber\\
&&\times  \exp\left(-\frac{s_{2,\infty }}{(1-s_{2,\infty })}z(1-t_2)(1-u_2)\right)\left\{ w_{2,2}^{\ast }\partial _{w_{2,2}^{\ast }} +\Big( \frac{1}{2}+ \frac{\lambda }{2}+ \frac{\omega }{2}\Big)\right\}\nonumber\\
&&\times  \int_{0}^{1} dt_1\;t_1^{-\frac{1}{2}+\frac{\lambda }{2}} \int_{0}^{1} du_1\;u_1^{\gamma -\frac{3}{2}+\frac{\lambda }{2}} 
\frac{1}{2\pi i}  \oint dv_1 \frac{\exp\left(-\frac{v_1}{(1-v_1)}w_{2,2}^{\ast }(1-t_1)(1-u_1)\right)}{(1-v_1)(v_1-s_1)}  \nonumber\\
&&\times \left\{\ddot{w}_{1,2}\partial _{\ddot{w}_{1,2}} +\Big(\frac{\lambda }{2}+ \frac{\omega }{2}\Big)\right\}
 \sum_{\beta _0=0}^{\infty }\frac{1}{\beta _0!} \left( \frac{s_{0,1}}{v_1}\right)^{\beta _0}\Bigg\{c_0 x^{\lambda} \sum_{i_0=0}^{\beta _0} \frac{(-\beta _0)_{i_0}}{(1+\frac{\lambda }{2})_{i_0}(\gamma +\frac{\lambda }{2})_{i_0}}\ddot{w}_{1,2}^{i_0}\Bigg\} \tilde{\varepsilon }^2 \label{eq:113}
\end{eqnarray}
By using Cauchy's integral formula, the contour integrand has poles at $v_1 =1$ or $s_1$,
and $s_1$ is only inside the unit circle. As we compute the residue there in (\ref{eq:113}) we obtain
\begin{eqnarray}
&&\sum_{\beta _0=0}^{\infty } \frac{s_0^{\beta _0}}{\beta _0!}\prod _{n=1}^{\infty } \left\{ \sum_{\beta _n=\beta _{n-1}}^{\infty } s_n^{\beta _n}\right\}y_2(x) \nonumber\\
&&= \prod_{k=2}^{\infty } \frac{1}{(1-s_{k,\infty })} \int_{0}^{1} dt_2\;t_2^{\frac{\lambda }{2}} \int_{0}^{1} du_2\;u_2^{\gamma -1+\frac{\lambda }{2}} \exp\left(-\frac{s_{2,\infty }}{(1-s_{2,\infty })}z(1-t_2)(1-u_2)\right) \nonumber\\
&&\times  \left( w_{2,2}^{\ast }\partial _{w_{2,2}^{\ast }} +\Big( \frac{1}{2}+ \frac{\lambda }{2}+ \frac{\omega }{2}\Big)\right)\nonumber\\
&&\times  \int_{0}^{1} dt_1\;t_1^{-\frac{1}{2}+\frac{\lambda }{2}} \int_{0}^{1} du_1\;u_1^{\gamma -\frac{3}{2}+\frac{\lambda }{2}} 
 \frac{\exp\left(-\frac{s_1}{(1-s_1)}w_{2,2}^{\ast }(1-t_1)(1-u_1)\right)}{(1-s_1)} \left( w_{1,2}^{\ast }\partial _{w_{1,2}^{\ast }} +\Big(\frac{\lambda }{2}+ \frac{\omega }{2}\Big)\right) \nonumber\\
&&\times  \sum_{\beta _0=0}^{\infty }\frac{s_0^{\beta _0}}{\beta _0!} \Bigg\{ c_0 x^{\lambda}\sum_{i_0=0}^{\beta _0} \frac{(-\beta _0)_{i_0}}{(1+\frac{\lambda }{2})_{i_0}(\gamma +\frac{\lambda }{2})_{i_0}}(w_{1,2}^{\ast })^{i_0}\Bigg\}\tilde{\varepsilon }^2 \label{eq:114}
\end{eqnarray}
where
\begin{eqnarray}
w_{1,2}^{\ast } &=& z s_{1,\infty } \prod _{l=1}^{2} t_l u_l\nonumber
\end{eqnarray}
Acting the summation operator $\displaystyle{ \sum_{\beta _0=0}^{\infty } \frac{s_0^{\beta _0}}{\beta _0!}\prod _{n=1}^{\infty } \left\{ \sum_{\beta _n=\beta _{n-1}}^{\infty } s_n^{\beta _n}\right\}}$ on (\ref{eq:105}),
\begin{eqnarray}
&&\sum_{\beta _0=0}^{\infty } \frac{s_0^{\beta _0}}{\beta _0!}\prod _{n=1}^{\infty } \left\{ \sum_{\beta _n=\beta _{n-1}}^{\infty } s_n^{\beta _n}\right\}y_3(x) \nonumber\\
&&= c_0 x^{\lambda} \tilde{\varepsilon }^3 \prod_{k=3}^{\infty } \frac{1}{(1-s_{k,\infty })} \int_{0}^{1} dt_3\;t_3^{\frac{1}{2}+\frac{\lambda }{2}} \int_{0}^{1} du_3\;u_3^{\gamma -\frac{1}{2}+\frac{\lambda }{2}} \exp\left(-\frac{s_{3,\infty }}{(1-s_{3,\infty })}z(1-t_3)(1-u_3)\right)\nonumber\\
&&\times  \left( w_{3,3}^{\ast }\partial _{w_{3,3}^{\ast }} +\Big( 1+ \frac{\lambda }{2}+ \frac{\omega }{2}\Big)\right)\nonumber\\
&&\times \int_{0}^{1} dt_2\;t_2^{\frac{\lambda }{2}} \int_{0}^{1} du_2\;u_2^{\gamma -1+\frac{\lambda }{2}} 
 \frac{\exp\left(-\frac{s_2}{(1-s_2)}w_{3,3}^{\ast }(1-t_2)(1-u_2)\right)}{(1-s_2)} 
\left( w_{2,3}^{\ast }\partial _{w_{2,3}^{\ast }} +\Big( \frac{1}{2}+ \frac{\lambda }{2}+ \frac{\omega }{2}\Big)\right)\nonumber\\
&&\times \int_{0}^{1} dt_1\;t_1^{-\frac{1}{2}+\frac{\lambda }{2}} \int_{0}^{1} du_1\;u_1^{\gamma -\frac{3}{2}+\frac{\lambda }{2}} 
 \frac{\exp\left(-\frac{s_1}{(1-s_1)}w_{2,3}^{\ast }(1-t_1)(1-u_1)\right)}{(1-s_1)}\left( w_{1,3}^{\ast }\partial _{w_{1,3}^{\ast }} +\Big( \frac{\lambda }{2}+ \frac{\omega }{2}\Big)\right)\nonumber\\
&&\times \sum_{\beta _0=0}^{\infty } \frac{s_0^{\beta _0}}{\beta _0!}\sum_{i_0=0}^{\beta _0}\frac{(-\beta _0)_{i_0}}{(1+\frac{\lambda }{2})_{i_0}(\gamma +\frac{\lambda }{2})_{i_0}}  (w_{1,3}^{\ast })^{i_0} \label{eq:115}
\end{eqnarray}
where
\begin{equation}
w_{3,3}^{\ast } = z s_{3,\infty } \prod _{l=3}^{3} t_l u_l \hspace{1cm} w_{2,3}^{\ast } = z s_{2,\infty } v_1\prod _{l=2}^{3} t_l u_l\hspace{1cm} w_{1,3}^{\ast } = z s_{1,\infty } v_1\prod _{l=1}^{3} t_l u_l\nonumber
\end{equation}
By repeating this process for all higher terms of integral forms of sub-summation $y_m(x)$ terms where $m > 3$, we obtain every  $\displaystyle{ \sum_{\beta _0=0}^{\infty } \frac{s_0^{\beta _0}}{\beta _0!}\prod _{n=1}^{\infty } \left\{ \sum_{\beta _n=\beta _{n-1}}^{\infty } s_n^{\beta _n}\right\}  y_m(x)}$ terms. 
Since we substitute (\ref{eq:106}), (\ref{eq:109}), (\ref{eq:114}), (\ref{eq:115}) and including all $\displaystyle{ \sum_{\beta _0=0}^{\infty } \frac{s_0^{\beta _0}}{\beta _0!}\prod _{n=1}^{\infty } \left\{ \sum_{\beta _n=\beta _{n-1}}^{\infty } s_n^{\beta _n}\right\}  y_m(x)}$ terms where $m > 3$ into (\ref{eq:106.1}), we obtain (\ref{eq:38})
\qed
\end{pot}
\begin{rmk}
The generating function for the GCH polynomial which makes $B_n$ term terminated of the first kind about $x=0$ as $\Omega = -2\mu (\beta  _i+\frac{i}{2})$ where $i,\beta _i= 0,1,2,\cdots$ is
\begin{eqnarray}
&&\sum_{\beta _0=0}^{\infty } \frac{s_0^{\beta _0}}{\beta _0!}\prod _{n=1}^{\infty } \left\{ \sum_{\beta _n=\beta _{n-1}}^{\infty } s_n^{\beta _n}\right\}QW_{\beta _i}\left( \beta _i=-\frac{\Omega }{2\mu }-\frac{i}{2}, \gamma =\frac{1}{2}(1+\nu );\; \tilde{\varepsilon }= -\frac{1}{2}\varepsilon x;\; z=-\frac{1}{2}\mu x^2 \right)\nonumber\\
&&=  \prod_{k=1}^{\infty } \frac{1}{(1-s_{k,\infty })} \displaystyle \mathbf{A} \left( s_{0,\infty } ;z\right)\nonumber\\
&&+ \left\{ \prod_{k=1}^{\infty } \frac{1}{(1-s_{k,\infty })} \int_{0}^{1} dt_1\;t_1^{-\frac{1}{2}} \int_{0}^{1} du_1\;u_1^{\gamma -\frac{3}{2}} \overleftrightarrow {\mathbf{\Gamma}}_1 \left(s_{1,\infty };t_1,u_1,z\right) \left( w_{1,1}^{\ast }\partial_{w_{1,1}^{\ast }} +\frac{\omega }{2}\right) \mathbf{A} \left(s_{0} ;w_{1,1}^{\ast }\right)\right\} \tilde{\varepsilon } \nonumber\\
&&+ \sum_{n=2}^{\infty } \Bigg\{ \prod_{k=n}^{\infty } \frac{1}{(1-s_{k,\infty })} \int_{0}^{1} dt_n\;t_n^{\frac{n}{2}-1} \int_{0}^{1} du_n \;u_n^{\gamma -2+\frac{n}{2}}   \overleftrightarrow {\mathbf{\Gamma}}_n \left(s_{n,\infty };t_n,u_n,z\right) \bigg( w_{n,n}^{\ast }\partial _{w_{n,n}^{\ast }}+ \frac{1}{2}\left( n-1 + \omega \right) \bigg)\nonumber\\
&&\times  \prod_{j=1}^{n-1} \Bigg\{ \int_{0}^{1} dt_{n-j}\;t_{n-j}^{\frac{1}{2}(n-j)-1} \int_{0}^{1} du_{n-j} \;u_{n-j}^{\gamma -2+\frac{1}{2}(n-j)} \overleftrightarrow {\mathbf{\Gamma}}_{n-j} \left(s_{n-j};t_{n-j},u_{n-j},w_{n-j+1,n}^{\ast }\right) \nonumber\\
&&\times \bigg( w_{n-j,n}^{\ast }\partial _{w_{n-j,n}^{\ast }}+ \frac{1}{2}\left( n-j-1 + \omega \right)\bigg)\Bigg\}
  \mathbf{A} \left(s_{0} ;w_{1,n}^{\ast }\right)  \Bigg\} \tilde{\varepsilon }^n 
 \label{eq:40}
\end{eqnarray}
where
\begin{equation}
\begin{cases} 
{ \displaystyle \overleftrightarrow {\mathbf{\Gamma}}_1 \left(s_{1,\infty };t_1,u_1,z\right)=  \exp \left(-\frac{s_{1,\infty }}{(1-s_{1,\infty })}z(1-t_1)(1-u_1)\right) }\cr
{ \displaystyle  \overleftrightarrow {\mathbf{\Gamma}}_n \left(s_{n,\infty };t_n,u_n,z\right) = \exp\left( -\frac{s_{n,\infty }}{(1-s_{n,\infty })}z(1-t_n)(1-u_n)\right)}\cr
{ \displaystyle \overleftrightarrow {\mathbf{\Gamma}}_{n-j} \left(s_{n-j};t_{n-j},u_{n-j},w_{n-j+1,n}^{\ast }\right) = \frac{\exp\left(-\frac{s_{n-j}}{(1-s_{n-j})}w_{n-j+1,n}^{\ast }(1-t_{n-j})(1-u_{n-j})\right)}{(1-s_{n-j})}}
\end{cases}\nonumber 
\end{equation}
and
\begin{equation}
\begin{cases} 
{ \displaystyle \mathbf{A} \left( s_{0,\infty } ;z\right)= (1-s_{0,\infty })^{-\gamma }\exp\left(-\frac{z s_{0,\infty }}{(1-s_{0,\infty })}\right)}\cr
{ \displaystyle  \mathbf{A} \left(s_{0} ;w_{1,1}^{\ast }\right) = (1-s_0)^{-\gamma }\exp\left(-\frac{w_{1,1}^{\ast }s_0}{(1-s_0)}\right)} \cr
{ \displaystyle \mathbf{A} \left(s_{0} ;w_{1,n}^{\ast }\right) = (1-s_0)^{-\gamma }\exp\left(-\frac{w_{1,n}^{\ast }s_0}{(1-s_0)}\right)}
\end{cases}\nonumber 
\end{equation}
\end{rmk}
\begin{pf}
The generating function for confluent Hypergeometric polynomial of the first kind is written by
\begin{equation}
\sum_{\beta _0=0}^{\infty } \frac{t^{\beta _0}}{\beta _0!} F_{\beta _0}(\gamma ;z)= (1-t)^{-\gamma } \exp\left(-\frac{zt}{(1-t)}\right) 
 \label{eq:39}
\end{equation}
Replace $t$ by $s_{0,\infty }$  in (\ref{eq:39}). 
\begin{equation}
\sum_{\beta _0=0}^{\infty } \frac{s_{0,\infty }^{\beta _0}}{\beta _0!} F_{\beta _0}(\gamma ;z)=  (1-s_{0,\infty })^{-\gamma } \exp\left(-\frac{zs_{0,\infty }}{(1-s_{0,\infty })}\right)  \label{eq:120}
\end{equation} 
Replace $t$ and $z$ by $s_0$ and $w_{1,1}^{\ast }$ in (\ref{eq:39}). 
\begin{equation}
\sum_{\beta _0=0}^{\infty } \frac{s_0^{\beta _0}}{\beta _0!} F_{\beta _0}(\gamma ;w_{1,1}^{\ast })=  (1-s_0)^{-\gamma } \exp\left(-\frac{w_{1,1}^{\ast }s_0 }{(1-s_0)}\right) \label{eq:121}
\end{equation}
Replace $t$ and $z$ by $s_0$ and $w_{1,n}^{\ast }$ in (\ref{eq:39}). 
\begin{equation}
\sum_{\beta _0=0}^{\infty } \frac{s_0^{\beta _0}}{\beta _0!} F_{\beta _0}(\gamma ;w_{1,n}^{\ast })=  (1-s_0)^{-\gamma } \exp\left(-\frac{w_{1,n}^{\ast }s_0 }{(1-s_0)}\right) \label{eq:122}
\end{equation} 
Put $c_0= \frac{\Gamma(\gamma +\beta _0) }{\Gamma (\gamma )}$ as $\lambda = 0$ in (\ref{eq:38}). And substitute (\ref{eq:120}), (\ref{eq:121}) and (\ref{eq:122}) into the new (\ref{eq:38}).
\qed
\end{pf}
\begin{rmk}
The generating function for the GCH polynomial which makes $B_n$ term terminated of the second kind about $x=0$ as $\Omega = -2\mu (\psi _i+1-\gamma +\frac{i}{2})$ where $i, \psi _i= 0,1,2,\cdots$ is 
\begin{eqnarray}
&&\sum_{\psi _0=0}^{\infty } \frac{s_0^{\psi _0}}{\psi _0!}\prod _{n=1}^{\infty } \left\{ \sum_{\psi _{n}=\psi _{n-1}}^{\infty } s_n^{\psi _n}\right\}RW_{\psi _i}\left( \psi _i=-\frac{\Omega }{2\mu }+\gamma -1-\frac{i}{2}, \gamma =\frac{1}{2}(1+\nu );\; \tilde{\varepsilon }= -\frac{1}{2}\varepsilon x;\; z=-\frac{1}{2}\mu x^2 \right) \nonumber\\
&&= z^{1-\gamma } \Bigg\{ \prod_{k=1}^{\infty } \frac{1}{(1-s_{k,\infty })} \mathbf{B} \left( s_{0,\infty } ;z\right)  \nonumber\\
 &&+ \left\{ \prod_{k=1}^{\infty } \frac{1}{(1-s_{k,\infty })} \int_{0}^{1} dt_1\;t_1^{\frac{1}{2}-\gamma } \int_{0}^{1} du_1\;u_1^{ -\frac{1}{2}}\overleftrightarrow {\mathbf{\Gamma}}_1 \left(s_{1,\infty };t_1,u_1,z\right)
  \left( w_{1,1}^{\ast }\partial_{w_{1,1}^{\ast }} +\left( \frac{\omega }{2}+1-\gamma \right)\right)\mathbf{B} \left( s_{0} ;w_{1,1}^{\ast }\right) \right\}\tilde{\varepsilon } \nonumber\\
&&+ \sum_{n=2}^{\infty } \Bigg\{ \prod_{k=n}^{\infty } \frac{1}{(1-s_{k,\infty })} \int_{0}^{1} dt_n\;t_n^{\frac{n}{2}-\gamma } \int_{0}^{1} du_n \;u_n^{-1+\frac{n}{2}} \overleftrightarrow {\mathbf{\Gamma}}_n \left(s_{n,\infty };t_n,u_n,z\right) \bigg( w_{n,n}^{\ast }\partial _{w_{n,n}^{\ast }}+ \Big(\frac{1}{2}(n+1+\omega )-\gamma  \Big) \bigg) \nonumber\\
&&\times  \prod_{j=1}^{n-1} \Bigg\{ \int_{0}^{1} dt_{n-j}\;t_{n-j}^{\frac{1}{2}(n-j)-\gamma } \int_{0}^{1} du_{n-j} \;u_{n-j}^{\frac{1}{2}(n-j)-1} \overleftrightarrow {\mathbf{\Gamma}}_{n-j} \left(s_{n-j};t_{n-j},u_{n-j},w_{n-j+1,n}^{\ast } \right)\nonumber\\
&&\times \bigg( w_{n-j,n}^{\ast }\partial _{w_{n-j,n}^{\ast }}+ \Big(\frac{1}{2}(n-j+1+\omega )-\gamma \Big)\bigg)\Bigg\} \mathbf{B} \left( s_{0} ;w_{1,n}^{\ast }\right) \Bigg\} \tilde{\varepsilon }^n \Bigg\}
  \label{eq:42}
\end{eqnarray}
where
\begin{equation}
\begin{cases} 
{ \displaystyle \overleftrightarrow {\mathbf{\Gamma}}_1 \left(s_{1,\infty };t_1,u_1,z\right)=  \exp \left(-\frac{s_{1,\infty }}{(1-s_{1,\infty })}z(1-t_1)(1-u_1)\right) }\cr
{ \displaystyle  \overleftrightarrow {\mathbf{\Gamma}}_n \left(s_{n,\infty };t_n,u_n,z\right) = \exp\left( -\frac{s_{n,\infty }}{(1-s_{n,\infty })}z(1-t_n)(1-u_n)\right)}\cr
{ \displaystyle \overleftrightarrow {\mathbf{\Gamma}}_{n-j} \left(s_{n-j};t_{n-j},u_{n-j},w_{n-j+1,n}^{\ast }\right) = \frac{\exp\left(-\frac{s_{n-j}}{(1-s_{n-j})}w_{n-j+1,n}^{\ast }(1-t_{n-j})(1-u_{n-j})\right)}{(1-s_{n-j})}}
\end{cases}\nonumber 
\end{equation}
and
\begin{equation}
\begin{cases} 
{ \displaystyle \mathbf{B} \left( s_{0,\infty } ;z\right)= (1-s_{0,\infty })^{ \gamma -2}\exp\left(-\frac{zs_{0,\infty }}{(1-s_{0,\infty })}\right)}\cr
{ \displaystyle  \mathbf{B} \left( s_{0} ;w_{1,1}^{\ast }\right) = (1-s_0)^{ \gamma -2} \exp\left(-\frac{w_{1,1}^{\ast }s_0}{(1-s_0)}\right)   } \cr
{ \displaystyle \mathbf{B} \left( s_{0} ;w_{1,n}^{\ast }\right) = (1-s_0)^{ \gamma -2}   \exp\left(-\frac{w_{1,n}^{\ast }s_0}{(1-s_0)} \right) }
\end{cases}\nonumber 
\end{equation}
\end{rmk}
\begin{pf}
The generating function for confluent hypergeometric polynomial of the second kind is written by
\begin{equation}
 \sum_{\psi _0=0}^{\infty } \frac{t^{\psi _0}}{\psi _0!} A_{\psi _0}(\gamma ;z)=  (1-t)^{ \gamma -2 } \exp\left(-\frac{zt}{(1-t)}\right)   
 \label{eq:41}
\end{equation}
Replace $t$ by $s_{0,\infty }$  in (\ref{eq:41}). 
\begin{equation}
\sum_{\psi _0=0}^{\infty } \frac{s_{0,\infty }^{\psi _0}}{\psi _0!} A_{\psi _0}(\gamma ;z) =  (1-s_{0,\infty })^{ \gamma -2}\exp\left(-\frac{zs_{0,\infty }}{(1-s_{0,\infty })}\right) \label{eq:123}
\end{equation} 
Replace $t$ and $z$ by $s_0$ and $w_{1,1}^{\ast }$ in (\ref{eq:41}). 
\begin{equation}
\sum_{\psi _0=0}^{\infty } \frac{s_0^{\psi _0}}{\psi _0!} A_{\psi _0}(\gamma ;w_{1,1}^{\ast }) =  (1-s_0)^{ \gamma -2} \exp\left(-\frac{w_{1,1}^{\ast }s_0}{(1-s_0)}\right) \label{eq:124}
\end{equation}
Replace $t$ and $z$ by $s_0$ and $w_{1,n}^{\ast }$ in (\ref{eq:41}). 
\begin{equation}
\sum_{\psi _0=0}^{\infty } \frac{s_0^{\psi _0}}{\psi _0!} A_{\psi _0}(\gamma ;w_{1,n}^{\ast }) =  (1-s_0)^{ \gamma -2}   \exp\left(-\frac{w_{1,n}^{\ast }s_0}{(1-s_0)} \right) \label{eq:125}
\end{equation} 
Put $c_0= \left( -\frac{1}{2}\mu \right)^{1-\gamma } \frac{\Gamma (\psi _0+2-\gamma )}{\Gamma (2-\gamma )}$ as $\lambda = 1-\nu = 2(1-\gamma )$  on (\ref{eq:38}) with replacing $\beta _0$, $\beta _{n-1}$ and $\beta _n$ by $\psi _0$, $\psi _{n-1}$ and $\psi _n$. Substitute (\ref{eq:123}), (\ref{eq:124}) and (\ref{eq:125}) into the new (\ref{eq:38}).
\qed
\end{pf}
\section{Application}
I show integral forms and generating functions for GCH polynomials of the first and second kinds in this paper. 
We can apply this new special function into many physics areas. I show three examples of GCH equation as follows:
\subsection{the rotating harmonic oscillator}
For example, there are quantum-mechanical systems whose radial Schr$\ddot{\mbox{o}}$dinger equation may be reduced to a Biconfluent Heun function\cite{Chou2012i,Leau1986,Mass1983}, namely the rotating harmonic oscillator and a class of confinement potentials. Its radial Schr$\ddot{\mbox{o}}$dinger equation is 
\begin{equation}
\Psi^{''}(r)+ \bigg\{ \frac{2\lambda _m+1}{2\omega } -\frac{(r-1)^2}{4\omega ^2}- \frac{l_m (l_m+1)}{r^2}\bigg\} \Psi(r) =0  
\label{eq:47}
\end{equation}
where $0\leq  r < \infty $, ¸$\lambda _m$ is the eigenvalue, $l_m$ is the rotational quantum number and $\omega$ is a
coupling parameter.
  
The wave function for the rotating harmonic oscillator is given by \cite{Chou2012i}
\begin{eqnarray}
\Psi(r) &=& N r^{l_m+1} \exp\left(-\frac{(r-1)^2}{2\omega }\right) QW_{\beta _i}\left( \beta _i=\frac{\lambda _m-l_m-1-i}{2}, \omega =l_m+1, \gamma =l_m+ \frac{3}{2}\right.\nonumber\\
&&\left.;\; \tilde{\varepsilon }= -\frac{r}{2\omega};\; z=\frac{r^2}{2\omega} \right)
\label{eq:48}
\end{eqnarray}
where
\begin{eqnarray}
&& QW_{\beta _i}\left( \beta _i=\frac{\lambda _m-l_m-1-i}{2} ,\omega =l_m+1, \gamma =l_m+ \frac{3}{2};\; \tilde{\varepsilon }= -\frac{r}{2\omega};\; z=\frac{r^2}{2\omega} \right)\nonumber\\
&&= \frac{\Gamma (\gamma +\beta _0)}{\Gamma (\gamma )} \Bigg\{\sum_{i_0=0}^{\beta _0 } \frac{(-\beta _0)_{i_0}}{(1)_{i_0}(\gamma)_{i_0}}z^{i_0}+   \Bigg\{\sum_{i_0=0}^{\beta _0 } \frac{(i_0+\frac{\omega }{2})}{(i_0+\frac{1}{2})(i_0-\frac{1}{2}+\gamma )} \frac{(-\beta  _0)_{i_0}}{(1)_{i_0}(\gamma)_{i_0}}\nonumber\\
&&\times  \sum_{i_1=i_0}^{\beta _1} \frac{(-\beta _1)_{i_1}(\frac{3}{2})_{i_0}(\gamma +\frac{1}{2})_{i_0}}{(-\beta _1)_{i_0}(\frac{3}{2})_{i_1}(\gamma +\frac{1}{2})_{i_1}} z^{i_1} \Bigg\} \tilde{\varepsilon }
+ \sum_{n=2}^{\infty } \Bigg\{ \sum_{i_0=0}^{\beta _0} \frac{(i_0+\frac{\omega }{2})}{(i_0+\frac{1}{2})(i_0-\frac{1}{2}+\gamma)} \frac{(-\beta _0)_{i_0}}{(1)_{i_0}(\gamma )_{i_0}} \nonumber\\
&&\times \prod _{k=1}^{n-1} \Bigg\{ \sum_{i_k=i_{k-1}}^{\beta _k} \frac{(i_k+\frac{\omega }{2}+\frac{k}{2})}{(i_k+\frac{1}{2}+\frac{k}{2})(i_k-\frac{1}{2}+\gamma + \frac{k}{2})} \frac{(-\beta _k)_{i_k}(1+\frac{k}{2})_{i_{k-1}}(\frac{k}{2}+\gamma )_{i_{k-1}}}{(-\beta _k)_{i_{k-1}}(1+\frac{k}{2})_{i_k}(\frac{k}{2}+\gamma )_{i_k}}\Bigg\} \nonumber\\
&&\times \sum_{i_n= i_{n-1}}^{\beta _n} \frac{(-\beta _n)_{i_n}(1+\frac{n}{2})_{i_{n-1}}(\frac{n}{2}+\gamma )_{i_{n-1}}}{(-\beta _n)_{i_{n-1}}(1+\frac{n}{2})_{i_n}(\frac{n}{2}+\gamma )_{i_n}} z^{i_n}\Bigg\} \tilde{\varepsilon }^n \Bigg\}
\label{eq:49}
\end{eqnarray}
N is normalized constant. GCH function with three recursive coefficients has infinite eigenvalues as we see (\ref{eq:49}) which is $\beta _i=\frac{\lambda _m-l_m-1-i}{2}$ where $i,\beta _i = 0,1,2,\cdots$.

We obtain the integral form of (\ref{eq:49}) from (\ref{eq:11}).
\begin{eqnarray}
&& QW_{\beta _i}\left( \beta _i=\frac{\lambda _m-l_m-1-i}{2} , \gamma =l_m+ \frac{3}{2};\; \tilde{\varepsilon }= -\frac{r}{2\omega};\; z=\frac{r^2}{2\omega} \right)\nonumber\\
&=& F_{\beta _0}(\gamma ;z) + \sum_{n=1}^{\infty } \Bigg\{\prod _{j=0}^{n-1} \Bigg\{ \int_{0}^{1} dt_{n-j}\;t_{n-j}^{\frac{1}{2}(n-j)-1} \int_{0}^{1} du_{n-j}\;u_{n-j}^{\gamma +\frac{1}{2}(n-j)-2} \nonumber\\
&&\times \frac{1}{2\pi i}  \oint dv_{n-j} \frac{\exp\left(-\frac{v_{n-j}}{(1-v_{n-j})}w_{n-j+1,n}(1-t_{n-j})(1-u_{n-j})\right)}{v_{n-j}^{\beta _{n-j}+1}(1-v_{n-j})}\nonumber\\
&&\times \left( w_{n-j,n}\partial _{w_{n-j,n}} +\frac{1}{2}\Big(n-j+l_m \Big)\right)\Bigg\} F_{\beta  _0}(\gamma ;w_{1,n})\Bigg\} \tilde{\varepsilon }^n  
\label{eq:50}
\end{eqnarray}
where
\begin{equation}w_{a,b}=
\begin{cases} \displaystyle {z\prod _{l=a}^{b} t_l u_l v_l }\cr
z \;\;\mbox{only}\;\mbox{if}\; a>b
\end{cases}
\label{eq:51}
\end{equation}
We can transform GCH function into all other well-known special functions having two recursive coefficients because a $_1F_1$ (or $F_{\beta _0}$) function recurs in each of sub-integral forms of GCH function in (\ref{eq:50}).
\subsection{Confinement potentials}
Following Chaudhuri and Mukherjee, there is the radial Schr$\ddot{\mbox{o}}$dinger equation.\cite{Chou2012i,Chau1983,Chau1984,Leau1986} 
\begin{equation}
\Psi^{''}(r)+ \bigg\{ \bigg( \frac{2\mu }{\hbar ^2} \bigg) \bigg( E+ \frac{a}{r} -br- cr^2\bigg) -\frac{l(l+1)}{r^2} \bigg\} \Psi(r) =0  
\label{eq:52}
\end{equation}
with E being the energy.

The wave function for confinement potentials is given by (see section 4.2 in Ref\cite{Chou2012i})
\begin{eqnarray}
\Psi(r) &=& N r^{l+1} \exp\left( -\frac{1}{2}r^2\alpha _F -\beta _F r \right) QW_{\beta _i}\left( \beta _i= \frac{1}{4 \alpha _F }\left( \beta _F^2 +\frac{2\mu }{\hbar ^2}E \right)-\frac{1}{2}\left(i+l+\frac{3}{2} \right) \right.\nonumber\\
&&,\left. \omega =-\frac{\mu a }{\hbar ^2 \beta _F}+ l+1, \gamma = l+ \frac{3}{2};\; \tilde{\varepsilon }= -\beta _F r;\; z=\alpha _F r^2\right) \label{eq:53}
\end{eqnarray}
where
\begin{eqnarray}
&& QW_{\beta _i}\left( \beta _i= \frac{1}{4 \alpha _F }\left( \beta _F^2 +\frac{2\mu }{\hbar ^2}E \right)-\frac{1}{2}\left(i+l+\frac{3}{2} \right),\omega =-\frac{\mu a }{\hbar ^2 \beta _F}+ l+1 \right.\nonumber\\
&&,\left. \gamma = l+ \frac{3}{2};\; \tilde{\varepsilon }= -\beta _F r;\; z=\alpha _F r^2\right)  \nonumber\\
&&= \frac{\Gamma (\gamma +\beta _0)}{\Gamma (\gamma )} \Bigg\{\sum_{i_0=0}^{\beta _0 } \frac{(-\beta _0)_{i_0}}{(1)_{i_0}(\gamma)_{i_0}}z^{i_0}+   \Bigg\{ \sum_{i_0=0}^{\beta _0 }\frac{(i_0+\frac{\omega }{2})}{(i_0+\frac{1}{2})(i_0-\frac{1}{2}+\gamma )} \frac{(-\beta  _0)_{i_0}}{(1)_{i_0}(\gamma)_{i_0}}\nonumber\\
&&\times  \sum_{i_1=i_0}^{\beta _1} \frac{(-\beta _1)_{i_1}(\frac{3}{2})_{i_0}(\gamma +\frac{1}{2})_{i_0}}{(-\beta _1)_{i_0}(\frac{3}{2})_{i_1}(\gamma +\frac{1}{2})_{i_1}} z^{i_1} \Bigg\} \tilde{\varepsilon }
+ \sum_{n=2}^{\infty } \Bigg\{ \sum_{i_0=0}^{\beta _0} \frac{(i_0+\frac{\omega }{2})}{(i_0+\frac{1}{2})(i_0-\frac{1}{2}+\gamma)} \frac{(-\beta _0)_{i_0}}{(1)_{i_0}(\gamma )_{i_0}} \nonumber\\
&&\times \prod _{k=1}^{n-1} \Bigg\{ \sum_{i_k=i_{k-1}}^{\beta _k} \frac{(i_k+\frac{\omega }{2}+\frac{k}{2})}{(i_k+\frac{1}{2}+\frac{k}{2})(i_k-\frac{1}{2}+\gamma + \frac{k}{2})} \frac{(-\beta _k)_{i_k}(1+\frac{k}{2})_{i_{k-1}}(\frac{k}{2}+\gamma )_{i_{k-1}}}{(-\beta _k)_{i_{k-1}}(1+\frac{k}{2})_{i_k}(\frac{k}{2}+\gamma )_{i_k}}\Bigg\} \nonumber\\
&&\times \sum_{i_n= i_{n-1}}^{\beta _n} \frac{(-\beta _n)_{i_n}(1+\frac{n}{2})_{i_{n-1}}(\frac{n}{2}+\gamma )_{i_{n-1}}}{(-\beta _n)_{i_{n-1}}(1+\frac{n}{2})_{i_n}(\frac{n}{2}+\gamma )_{i_n}} z^{i_n}\Bigg\} \tilde{\varepsilon }^n \Bigg\}
\label{eq:54}
\end{eqnarray}
N is normalized constant. Energy $E$ is given by
\begin{equation}
E= \frac{\hbar ^2}{2\mu } \left(  4 \alpha _F \left( \beta _i+\frac{i+l+\frac{3}{2}}{2} \right) -\beta _F^2\right)\hspace{1cm} \mathrm{where}\; i,\beta _i= 0,1,2,\cdots  
\label{eq:55}
\end{equation}
GCH function with three recursive coefficients has infinite eigenvalues as we see (\ref{eq:54}). 
We obtain the integral form of (\ref{eq:54}) from (\ref{eq:11}).
\begin{eqnarray}
&& QW_{\beta _i}\left( \beta _i= \frac{1}{4 \alpha _F }\left( \beta _F^2 +\frac{2\mu }{\hbar ^2}E \right)-\frac{1}{2}\left(i+l+\frac{3}{2} \right),\omega =-\frac{\mu a }{\hbar ^2 \beta _F}+ l+1  \right.\nonumber\\
&&,\left. \gamma = l+ \frac{3}{2};\; \tilde{\varepsilon }= -\beta _F r;\; z=\alpha _F r^2\right) \nonumber\\
&=& F_{\beta _0}(\gamma ;z) + \sum_{n=1}^{\infty } \Bigg\{\prod _{j=0}^{n-1} \Bigg\{ \int_{0}^{1} dt_{n-j}\;t_{n-j}^{\frac{1}{2}(n-j)-1} \int_{0}^{1} du_{n-j}\;u_{n-j}^{\gamma +\frac{1}{2}(n-j)-2} \nonumber\\
&&\times \frac{1}{2\pi i}  \oint dv_{n-j} \frac{\exp\left(-\frac{v_{n-j}}{(1-v_{n-j})}w_{n-j+1,n}(1-t_{n-j})(1-u_{n-j})\right)}{v_{n-j}^{\beta _{n-j}+1}(1-v_{n-j})}\nonumber\\
&&\times \left( w_{n-j,n}\partial _{w_{n-j,n}} +\frac{1}{2}\Big(n-j+l-\frac{a}{2\beta _F}\Big)\right) \Bigg\} F_{\beta  _0}(\gamma ;w_{1,n})\Bigg\}  \tilde{\varepsilon }^n  
\label{eq:56}
\end{eqnarray}
where
\begin{equation}w_{a,b}=
\begin{cases} \displaystyle {z\prod _{l=a}^{b} t_l u_l v_l }\cr
z \;\;\mbox{only}\;\mbox{if}\; a>b
\end{cases}
\label{eq:57}
\end{equation}
Again, $_1F_1$ (or $F_{\beta _0}$) function recurs in each of sub-integral forms in (\ref{eq:56}). 
\subsection{Two interacting electrons in a uniform magnetic field and a parabolic potential}
There are three identical charged particles on a plane under a perpendicular magnetic field and interacting through Coulomb repulsion.\cite{Ralk2002} Its solution is also Biconflent Heun function which is the special case of GCH function. 
In ``Two interacting electrons in a uniform magnetic field and a parabolic potential: The general closed-form solution''\cite{Kand2005}, the author consider a system of two interacting electrons of mass $m^{\ast}$ and charge $e$ in two space dimensions subjected to both a uniform magnetic field along the direction perpendicular to the plane and an external parabolic potential. 
Its Hamiltonian is given by
\begin{equation}
H = \sum_{j=1}^{2} \Bigg\{ \frac{1}{2 m^{\ast}}\bigg[ \textbf{p}(\textbf{r}_j)+ \frac{e}{c}\textbf{A}(\textbf{r}_j) \bigg]^2 + U(|\textbf{r}_j|)\Bigg\} + \frac{e^2}{\epsilon _{\infty }|\textbf{r}_1 -\textbf{r}_2|}
\label{eq:58}
\end{equation}
where $U(|\textbf{r}_j|)=1/2 m^{\ast}\omega ^2\textbf{r}_j^2$ is the single particle confinement potential, and $\textbf{A}(\textbf{r}_j) $ is the vector potential of the magnetic field. By introducing the relative and center-of-mass coordinates,$\textbf{r}= |\textbf{r}_1 -\textbf{r}_2|$ and $\textbf{R}= (\textbf{r}_1 +\textbf{r}_2)/2$, respectively, (\ref{eq:58}) can be decoupled as the sum of two single particle Hamiltonians. by setting $\Psi (\textbf{r}) = e^{im \varphi  } R(r) $ , the equation for the radial part of the relative motion in the cylindrical coordinates is
\begin{equation}
\frac{d^2 R}{d \rho ^2}+ \frac{1}{\rho }\frac{d R}{d \rho} - \frac{\sigma ^2 m^2}{\rho ^2} R - \rho ^2 R -\frac{u}{\rho } R +\tilde{\epsilon }R=0
\label{eq:59}
\end{equation}
where 
\begin{equation}
\begin{split}
& \tilde{\epsilon }  = (2E_r-m\hbar \omega _c)/\hbar \omega \\ &  u= 2\mu e^2/\epsilon _{\infty }\hbar ^2\tilde{\gamma} \\ &  \rho =\tilde{\gamma}r  \\
& \tilde{\gamma}^2= \mu \omega /\hbar  
\end{split}\label{eq:60}   
\end{equation}
Put $R(\rho )= \rho ^{\sigma |m|}e^{-\rho ^2/2}F(\rho ) $ in (\ref{eq:59}).
\begin{equation}
\rho F^{''} +(a-2\rho ^2)F^{'}+(d\rho -u)F=0
\label{eq:60}
\end{equation}
where
\begin{equation}
\begin{split}
& a= 2\sigma |m|+1 \\ & d= \tilde{\epsilon }-(2\sigma |m|+1) 
\end{split}\label{eq:61}   
\end{equation}
(\ref{eq:60}) is generally called the Biconfluent Heun equation in canonical form. 
If we compare (\ref{eq:60}) with (\ref{eq:1a}), all coefficients on the above are correspondent to the following way.
\begin{equation}
\begin{split}
& \mu  \longleftrightarrow   -2 \\ & \varepsilon  \longleftrightarrow  0  \\ & \nu  \longleftrightarrow  a= 2\sigma |m|+1 \\
& \Omega  \longleftrightarrow  d= \tilde{\epsilon }-(2\sigma |m|+1)  \\ & \omega  \longleftrightarrow \frac{-u}{0} \\ & x  \longleftrightarrow \rho  
\end{split}\label{eq:62}   
\end{equation}
Let's investigate function $\Psi(r)$ as $n$ and $r$ go to infinity. I assume that $F(\rho )$ is infinite series in (\ref{eq:60}).
In Ref.\cite{Chou2012i}, the asymptomatic behavior of ${\displaystyle \lim_{n\gg 1}y(x)}$ for the case of $|\varepsilon| \ll |\mu |$ is 
\begin{equation}
\lim_{n\gg 1}y(x) = 1+ \sqrt{ -\frac{\pi }{2}\mu x^2} \mbox{Erf}\bigg(\sqrt{-\frac{1}{2}\mu x^2}\bigg) \exp\left( -\frac{1}{2}\mu x^2\right)  \hspace{1cm}\mbox{where}\;-\infty <x< \infty 
\label{eq:63}
\end{equation}
Replacing $y(x)$ and $\mu $ by $F(\rho )$ and $-2$ in (\ref{eq:63}). Take the new (\ref{eq:63}) into $\Psi (\textbf{r}) = e^{im \varphi  } R(\rho )= \rho ^{\sigma |m|}e^{-\rho ^2/2}F(\rho )e^{im \varphi}$ putting $\rho =\tilde{\gamma }r $.
\begin{equation}
\lim_{n\gg 1}\Psi(r) \approx (\tilde{\gamma }r)^{\sigma |m|} \exp\left(-\frac{(\tilde{\gamma }r)^2}{2}\right)\left\{ 1+ \tilde{\gamma } \sqrt{\pi}\;\mbox{Erf}(\tilde{\gamma }r)\; r \exp\left(\tilde{\gamma }^2 r^2\right) \right\}\exp\left(im \varphi\right) 
\label{eq:64}
\end{equation}
 In (\ref{eq:64}) if $r\rightarrow \infty $, then $\displaystyle {\lim_{n\gg 1}\Psi(r)\rightarrow \infty }$. It is unacceptable that wave function $\Psi(r)$ is divergent as $r$ goes to infinity in the quantum mechanical point of view. Therefore the function $F(\rho )$ must to be polynomial in (\ref{eq:60}) in order to make the wave function $\Psi(r)$ being convergent even if $r$ goes to infinity.
 
In Ref.\cite{Chou2012i}, the Frobenius solutions of GCH polynomial which makes $B_n$ term terminated of the first and second kinds are given by
\begin{eqnarray}
 y(x)&=& QW_{\beta _i}\left( \beta _i=-\frac{\Omega }{2\mu }-\frac{i}{2} , \omega,  \gamma =\frac{1}{2}(1+\nu );\; \tilde{\varepsilon }= -\frac{1}{2}\varepsilon x;\; z=-\frac{1}{2}\mu x^2 \right) \nonumber\\
&&= \frac{\Gamma (\gamma +\beta _0)}{\Gamma (\gamma )} \Bigg\{\sum_{i_0=0}^{\beta _0 } \frac{(-\beta _0)_{i_0}}{(1)_{i_0}(\gamma)_{i_0}}z^{i_0}+  \Bigg\{ \sum_{i_0=0}^{\beta _0 } \frac{(i_0+\frac{\omega }{2})}{(i_0+\frac{1}{2})(i_0-\frac{1}{2}+\gamma )} \frac{(-\beta  _0)_{i_0}}{(1)_{i_0}(\gamma)_{i_0}}\nonumber\\
&&\times  \sum_{i_1=i_0}^{\beta _1} \frac{(-\beta _1)_{i_1}(\frac{3}{2})_{i_0}(\gamma +\frac{1}{2})_{i_0}}{(-\beta _1)_{i_0}(\frac{3}{2})_{i_1}(\gamma +\frac{1}{2})_{i_1}} z^{i_1} \Bigg\}\tilde{\varepsilon }
+ \sum_{n=2}^{\infty } \Bigg\{ \sum_{i_0=0}^{\beta _0}  \frac{(i_0+\frac{\omega }{2})}{(i_0+\frac{1}{2})(i_0-\frac{1}{2}+\gamma)} \frac{(-\beta _0)_{i_0}}{(1)_{i_0}(\gamma )_{i_0}} \nonumber\\
&&\times \prod _{k=1}^{n-1} \Bigg\{ \sum_{i_k=i_{k-1}}^{\beta _k} \frac{(i_k+\frac{\omega }{2}+\frac{k}{2})}{(i_k+\frac{1}{2}+\frac{k}{2})(i_k-\frac{1}{2}+\gamma + \frac{k}{2})} \frac{(-\beta _k)_{i_k}(1+\frac{k}{2})_{i_{k-1}}(\frac{k}{2}+\gamma )_{i_{k-1}}}{(-\beta _k)_{i_{k-1}}(1+\frac{k}{2})_{i_k}(\frac{k}{2}+\gamma )_{i_k}}\Bigg\} \nonumber\\
&&\times \sum_{i_n= i_{n-1}}^{\beta _n} \frac{(-\beta _n)_{i_n}(1+\frac{n}{2})_{i_{n-1}}(\frac{n}{2}+\gamma )_{i_{n-1}}}{(-\beta _n)_{i_{n-1}}(1+\frac{n}{2})_{i_n}(\frac{n}{2}+\gamma )_{i_n}} z^{i_n} \Bigg\} \tilde{\varepsilon }^n \Bigg\}
\label{eq:65}
\end{eqnarray}
\begin{eqnarray}
 y(x)&=& RW_{\psi _i}\left( \psi _i=-\frac{\Omega }{2\mu }+\gamma -1-\frac{i}{2}, \omega, \gamma =\frac{1}{2}(1+\nu );\; \tilde{\varepsilon }= -\frac{1}{2}\varepsilon x;\; z=-\frac{1}{2}\mu x^2 \right) \nonumber\\
&&= z^{1-\gamma }\frac{\Gamma (\psi _0+2-\gamma )}{\Gamma (2-\gamma )} \Bigg\{\sum_{i_0=0}^{\psi _0} \frac{(-\psi _0)_{i_0}}{(1)_{i_0}(2-\gamma)_{i_0}}z^{i_0}\nonumber\\
&&+  \Bigg\{\sum_{i_0=0}^{\psi _0} \frac{(i_0+1-\gamma +\frac{\omega }{2})}{(i_0+\frac{1}{2})(i_0+\frac{3}{2}-\gamma )} \frac{(-\psi _0)_{i_0}}{(1)_{i_0}(2-\gamma)_{i_0}}
  \sum_{i_1=i_0}^{\psi _1} \frac{(-\psi _1)_{i_1}(\frac{3}{2})_{i_0}(\frac{5}{2}-\gamma )_{i_0}}{(-\psi _1)_{i_0}(\frac{3}{2})_{i_1}(\frac{5}{2}-\gamma )_{i_1}} z^{i_1} \Bigg\}  \tilde{\varepsilon } \nonumber\\
&&+ \sum_{n=2}^{\infty } \Bigg\{ \sum_{i_0=0}^{\psi _0} \frac{(i_0+1-\gamma +\frac{\omega }{2})}{(i_0+\frac{1}{2})(i_0+\frac{3}{2}-\gamma)} \frac{(-\psi _0)_{i_0}}{(1)_{i_0}(2-\gamma )_{i_0}} \nonumber\\
&&\times \prod _{k=1}^{n-1} \Bigg\{ \sum_{i_k=i_{k-1}}^{\psi _k} \frac{(i_k+1-\gamma +\frac{\omega }{2}+\frac{k}{2})}{(i_k+\frac{1}{2}+\frac{k}{2})(i_k+\frac{3}{2}-\gamma + \frac{k}{2})} \frac{(-\psi _k)_{i_k}(1+\frac{k}{2})_{i_{k-1}}(2-\gamma +\frac{k}{2})_{i_{k-1}}}{(-\psi _k)_{i_{k-1}}(1+\frac{k}{2})_{i_k}(2-\gamma +\frac{k}{2})_{i_k}}\Bigg\} \nonumber\\
&&\times \sum_{i_n= i_{n-1}}^{\psi _n} \frac{(-\psi _n)_{i_n}(1+\frac{n}{2})_{i_{n-1}}(2-\gamma +\frac{n}{2})_{i_{n-1}}}{(-\psi _n)_{i_{n-1}}(1+\frac{n}{2})_{i_n}(2-\gamma +\frac{n}{2})_{i_n}} z^{i_n}\Bigg\}   \tilde{\varepsilon }^n \Bigg\}
\label{eq:66}
\end{eqnarray}
Since (\ref{eq:62}) is put in (\ref{eq:65}) and (\ref{eq:66}), $RW_{\psi _i}\left( \psi _i, \omega, \gamma;\; \tilde{\varepsilon };\; z\right)\rightarrow \infty $ as $r\rightarrow 0$ because of $ \gamma =\sigma |m|+1$. And $QW_{\beta _i}\left( \beta _i,\omega, \gamma;\; \tilde{\varepsilon };\; z\right)\rightarrow 0$ as $r\rightarrow 0$. So I choose $QW_{\beta _i}\left( \beta _i,\omega, \gamma;\; \tilde{\varepsilon };\; z\right)$ as an eigenfunction for (\ref{eq:60}). Apply (\ref{eq:62}) into (\ref{eq:65}) with replacing $y(x)$ by $F(\rho=\tilde{\gamma}r)$.
\begin{eqnarray}
 F(\rho) &=& QW_{\beta _i}\left( \beta _i=\frac{d}{4}-\frac{i}{2}, \gamma=\sigma |m|+1;\; \tilde{\varepsilon }= \frac{\mu e^2}{2\epsilon _{\infty }\hbar ^2}r ;\; z=\frac{\mu \omega }{\hbar}r^2 \right) \nonumber\\
&=& \frac{\Gamma (\gamma+\beta _0)}{\Gamma (\gamma)} \Bigg\{\sum_{i_0=0}^{\beta _0 } \frac{(-\beta _0)_{i_0}}{(1)_{i_0}(\gamma)_{i_0}}z^{i_0}+  \Bigg\{ \sum_{i_0=0}^{\beta _0 }\frac{1}{(i_0+\frac{1}{2})(i_0-\frac{1}{2}+\gamma)} \frac{(-\beta  _0)_{i_0}}{(1)_{i_0}(\gamma)_{i_0}}\nonumber\\
&&\times  \sum_{i_1=i_0}^{\beta _1} \frac{(-\beta _1)_{i_1}(\frac{3}{2})_{i_0}(\gamma +\frac{1}{2})_{i_0}}{(-\beta _1)_{i_0}(\frac{3}{2})_{i_1}(\gamma +\frac{1}{2})_{i_1}} z^{i_1}  \Bigg\}\tilde{\varepsilon } 
+ \sum_{n=2}^{\infty } \Bigg\{ \sum_{i_0=0}^{\beta _0} \frac{1}{(i_0+\frac{1}{2})(i_0-\frac{1}{2}+\gamma)} \frac{(-\beta _0)_{i_0}}{(1)_{i_0}(\gamma)_{i_0}} \nonumber\\
&&\times \prod _{k=1}^{n-1} \Bigg\{ \sum_{i_k=i_{k-1}}^{\beta _k} \frac{1}{(i_k+\frac{1}{2}+\frac{k}{2})(i_k-\frac{1}{2}+\gamma+ \frac{k}{2})} \frac{(-\beta _k)_{i_k}(1+\frac{k}{2})_{i_{k-1}}(\frac{k}{2}+\gamma)_{i_{k-1}}}{(-\beta _k)_{i_{k-1}}(1+\frac{k}{2})_{i_k}(\frac{k}{2}+\bar{\gamma})_{i_k}}\Bigg\} \nonumber\\
&&\times \sum_{i_n= i_{n-1}}^{\beta _n} \frac{(-\beta _n)_{i_n}(1+\frac{n}{2})_{i_{n-1}}(\frac{n}{2}+\gamma)_{i_{n-1}}}{(-\beta _n)_{i_{n-1}}(1+\frac{n}{2})_{i_n}(\frac{n}{2}+\gamma)_{i_n}} z^{i_n} \Bigg\} \tilde{\varepsilon }^n \Bigg\}
\label{eq:67}
\end{eqnarray}
Put (\ref{eq:67}) in $\Psi (\textbf{r}) = \rho ^{\sigma |m|}e^{-\rho ^2/2}F(\rho )e^{im \varphi}$ where $\rho =\tilde{\gamma}r= \sqrt{\frac{\mu \omega }{\hbar}}r$. The wave function in a uniform magnetic field and a parabolic potential is given by
\begin{equation}
\Psi(r) = N \left(\frac{\mu \omega }{\hbar}r^2\right)^{\frac{\sigma |m|}{2}}\exp\left(-\frac{\mu \omega }{2\hbar}r^2\right) QW_{\beta _i}\left( \beta _i=\frac{d}{4}-\frac{i}{2}, \gamma=\sigma |m|+1;\; \tilde{\varepsilon }= \frac{\mu e^2}{2\epsilon _{\infty }\hbar ^2}r ;\; z=\frac{\mu \omega }{\hbar}r^2 \right) e^{im \varphi}
\label{eq:68}
\end{equation}
N is normalized constant. Energy $E_r$ is given by
\begin{equation}
E_r= \hbar \omega \left(2\beta _i+i+\sigma |m|+\frac{1}{2}\right) +\frac{1}{2}m\hbar \omega _c
\hspace{1cm} \mathrm{where}\; i,\beta _i= 0,1,2,\cdots
\label{eq:69}
\end{equation}
GCH function with three recursive coefficients has infinite eigenvalues as we see (\ref{eq:67}). 
We obtain the integral form of (\ref{eq:67}) from (\ref{eq:11}).
\begin{eqnarray}
F(\rho) &=& QW_{\beta _i}\left( \beta _i=\frac{d}{4}-\frac{i}{2}, \gamma=\sigma |m|+1;\; \tilde{\varepsilon }= \frac{\mu e^2}{2\epsilon _{\infty }\hbar ^2}r ;\; z=\frac{\mu \omega }{\hbar}r^2 \right)\nonumber\\
&=& F_{\beta _0}(\gamma ;z) + \sum_{n=1}^{\infty } \Bigg\{\prod _{j=0}^{n-1} \Bigg\{ \int_{0}^{1} dt_{n-j}\;t_{n-j}^{\frac{1}{2}(n-j)-1} \int_{0}^{1} du_{n-j}\;u_{n-j}^{\gamma +\frac{1}{2}(n-j)-2} \label{eq:70}\\
&&\times \frac{1}{2\pi i}  \oint dv_{n-j} \frac{\exp\left(-\frac{v_{n-j}}{(1-v_{n-j})}w_{n-j+1,n}(1-t_{n-j})(1-u_{n-j})\right)}{v_{n-j}^{\beta _{n-j}+1}(1-v_{n-j})}
 w_{n-j,n}\partial _{w_{n-j,n}} \Bigg\} F_{\beta  _0}(\gamma ;w_{1,n})\Bigg\} \tilde{\varepsilon }^n  
\nonumber
\end{eqnarray}
where
\begin{equation}w_{a,b}=
\begin{cases} \displaystyle {z\prod _{l=a}^{b} t_l u_l v_l }\cr
z \;\;\mbox{only}\;\mbox{if}\; a>b
\end{cases}
\label{eq:71}
\end{equation}
$_1F_1$ (or $F_{\beta _0}$) function again recurs in each of sub-integral forms in (\ref{eq:70}). 
\section{Conclusion}
In Ref.\cite{Chou2012i} I show the power series expansion in closed forms and its asymptotic behaviors of the GCH equation for infinite series and polynomial. In this paper I construct integral forms of GCH equation for infinite series and polynomial which makes $B_n$ term terminated including all higher terms of $A_n$'s. Indeed, the generating functions for GCH polynomial are derived in mathematical rigour. And I show how to derive the power series expansions in closed forms and its integral forms of analytic wave functions and its eigenvalues in four examples: (1) Schr$\ddot{\mbox{o}}$dinger equation with the rotating harmonic oscillator and a class of confinement potentials, (2) The spin free Hamiltonian involving only scalar potential for the $q-\bar{q}$ system  \cite{Chou2012i}, (3) The radial Schr$\ddot{\mbox{o}}$dinger equation with Confinement potentials, (4) Two interacting electrons in a uniform magnetic field and a parabolic potential. 

In general, most of wave-functions in physics are quantized with specific eigenvalues. As we see three examples on the above, all solutions are quantized with certain eigenvalues and its analytic wave-functions have polynomial expansions in closed forms. There are infinite eigenvalues because its differential equations have three recursive coefficients\cite{Chou2012a,Chou2012b,Chou2012i}. In contrast most of well-known wave functions only have one eigenvalue because its differential equations have two recursive coefficients.

We can express representations in closed form integrals in an easy way since we have power series expansions with Pochhammer symbols in numerators and denominators. We can transform any special functions, having three term recursive relation, into all other well-known special functions with two recursive coefficients because a $_1F_1$ (GCH function) or $_2F_1$ (Mathieu, Lame, Heun, functions\cite{Chou2012d,Chou2012e,Chou2012f,Chou2012g}) function recurs in each of sub-integral forms of them. It means all analytic solutions in the three-term recurrence can be described as hypergoemetric function: understanding the connection between other special functions is important in the mathematical and physical points of views as we all know.

The analytic integral form of the GCH function with three recursive coefficients are derived from power series expansion in closed forms of the GCH function. And generating function for the GCH polynomial is constructed from the integral form of the GCH polynomial. The generating function is really helpful in order to derive orthogonal relations, recursion relations and expectation values of any physical quantities. For the case of hydrogen-like atoms, the normalized wave function is derived from the generating function for associated Laguerre polynomials. And expectation values of physical quantities such as position and momentum are constructed by applying the recursive relation of associated Laguerre polynomials.
\vspace{5mm}

\section{Series ``Special functions and three term recurrence formula (3TRF)''} 

This paper is 10th out of 10.
\vspace{3mm}

1. ``Approximative solution of the spin free Hamiltonian involving only scalar potential for the $q-\bar{q}$ system'' \cite{Chou2012a} - In order to solve the spin-free Hamiltonian with light quark masses we are led to develop a totally new kind of special function theory in mathematics that generalize all existing theories of confluent hypergeometric types. We call it the Grand Confluent Hypergeometric Function. Our new solution produces previously unknown extra hidden quantum numbers relevant for description of supersymmetry and for generating new mass formulas.
\vspace{3mm}

2. ``Generalization of the three-term recurrence formula and its applications'' \cite{Chou2012b} - Generalize three term recurrence formula in linear differential equation.  Obtain the exact solution of the three term recurrence for polynomials and infinite series.
\vspace{3mm}

3. ``The analytic solution for the power series expansion of Heun function'' \cite{Chou2012c} -  Apply three term recurrence formula to the power series expansion in closed forms of Heun function (infinite series and polynomials) including all higher terms of $A_n$'s.
\vspace{3mm}

4. ``Asymptotic behavior of Heun function and its integral formalism'', \cite{Chou2012d} - Apply three term recurrence formula, derive the integral formalism, and analyze the asymptotic behavior of Heun function (including all higher terms of $A_n$'s). 
\vspace{3mm}

5. ``The power series expansion of Mathieu function and its integral formalism'', \cite{Chou2012e} - Apply three term recurrence formula, analyze the power series expansion of Mathieu function and its integral forms.  
\vspace{3mm}

6. ``Lame equation in the algebraic form'' \cite{Chou2012f} - Applying three term recurrence formula, analyze the power series expansion of Lame function in the algebraic form and its integral forms.
\vspace{3mm}

7. ``Power series and integral forms of Lame equation in   Weierstrass's form and its asymptotic behaviors'' \cite{Chou2012g} - Applying three term recurrence formula, derive the power series expansion of Lame function in   Weierstrass's form and its integral forms. 
\vspace{3mm}

8. ``The generating functions of Lame equation in   Weierstrass's form'' \cite{Chou2012h} - Derive the generating functions of Lame function in   Weierstrass's form (including all higher terms of $A_n$'s).  Apply integral forms of Lame functions in   Weierstrass's form.
\vspace{3mm}

9. ``Analytic solution for grand confluent hypergeometric function'' \cite{Chou2012i} - Apply three term recurrence formula, and formulate the exact analytic solution of grand confluent hypergeometric function (including all higher terms of $A_n$'s). Replacing $\mu $ and $\varepsilon \omega $ by 1 and $-q$, transforms the grand confluent hypergeometric function into Biconfluent Heun function.
\vspace{3mm}

10. ``The integral formalism and the generating function of grand confluent hypergeometric function'' \cite{Chou2012j} - Apply three term recurrence formula, and construct an integral formalism and a generating function of grand confluent hypergeometric function (including all higher terms of $A_n$'s). 
\section*{Acknowledgment}
I thank Bogdan Nicolescu. The discussions I had with him on number theory was of great joy.  
\vspace{3mm}

\bibliographystyle{model1a-num-names}
\bibliography{<your-bib-database>}

\end{document}